\documentstyle[aps,epsfig]{revtex}

\newcommand{\be}{\begin{equation}}
\newcommand{\ee}{\end{equation}}
\newcommand{\bea}{\begin{eqnarray}}
\newcommand{\eea}{\end{eqnarray}}
\newcommand{\inli}{\int\limits}

\newcommand{\bk}{\bf k}
\newcommand{\bkp}{\bf k_\perp}

\newcommand{\al}{\alpha}
\newcommand{\bet}{\beta}

\def\la{\mathrel{\mathpalette\fun <}}

\def\fun#1#2{\lower3.6pt\vbox{\baselineskip0pt\lineskip.9pt
\ialign{$\mathsurround=0pt#1\hfil##\hfil$\crcr#2\crcr\sim\crcr}}}

\begin{document}

\title{Determination of quark--antiquark component of the photon wave
function  for $u,d,s$-quarks }

\author{A.V. Anisovich, V.V. Anisovich, L.G. Dakhno, V.A. Nikonov,\\
 A.V.  Sarantsev}
\date{}
\maketitle

\begin{abstract}

Based on the data for the transitions
 $\pi^0, \eta,\eta'\to \gamma\gamma^*(Q^2)$
and reactions of the $ e^+e^-$-annihilations,
$ e^+e^-\to \rho^0,\omega,\phi$ and $ e^+e^-\to hadrons$ at
$1<E_{e^+e^-}<3.7$ GeV, we determine the light-quark components of
the photon wave function $\gamma^*(Q^2)\to q\bar q$  ($q=u,d,s$) for
the region $0\la Q^2 \la 1$ (GeV/{\it c})$^2$.

\end{abstract}

\section{Introduction}

 In the search for exotic states one needs to find out the
quark--gluon content of  mesons and  establish meson systematics.
The meson radiative decay is  a powerful tool for qualitative
evaluation of the quark--antiquark components.
The study of the two-photon transitions such as $meson\to
\gamma\gamma$ and, more generally, $meson\to
\gamma^*(Q_1^2)\gamma^*(Q_2^2)$,  looks as a promising way to reveal
the quark--antiquark content of  mesons.

Experimental data accumulated by the collaborations L3 \cite{L3,L3new},
ARGUS \cite{ARGUS},
CELLO \cite{CELLO}, TRC/$2\gamma$ \cite{TPC}, CLEO \cite{CLEO},
Mark II \cite{Mark}, Crystal Ball \cite{Crystal}, and others
make it obvious that the calculation of the processes $meson\to
\gamma^*(Q^2_1)\gamma^*(Q^2_2)$ is up to date. To make this reaction
informative as concern the meson quark--gluon content
one needs a reliably determined wave function of the photon at
$0\la Q^2 \la 1$ (GeV/{\it c})$^2$. Conventionally, one may
consider two pieces of the photon wave function: soft and hard ones.
Hard component relates to the point-like vertex $\gamma\to q\bar q$, it
is responsible for the production of quark--antiquark pair
at high virtuality.
At large energy of the $e^+e^-$ system, the ratio of cross
sections $R=\sigma(e^+e^-\to hadrons)/\sigma(e^+e^-\to \mu^+\mu^-)$ is
determined by the hard component of photon wave function, while soft
component is responsible for the production of low-energy
quark--antiquark vector states such as
 $\rho^0$, $\omega$, $\phi(1020)$, and their excitions.

Evaluation of the photon wave function for the
$\gamma^*(Q^2)\to u\bar u,d\bar d, s\bar s$ transitions
was carried out in \cite{PR97}, on the basis of
data of the CLEO Collaboration on the $Q^2$-dependent
transition form factors $\pi^0\to \gamma\gamma^*(Q^2)$,
$\eta\to \gamma\gamma^*(Q^2)$, and $\eta'\to \gamma\gamma^*(Q^2)$, see
\cite{CLEO} and references therein.
The goal of the present paper
is,  by adding the information on the process $e^+e^-\to
hadrons$, to define more precisely the wave function
$\gamma^*(Q^2)\to u\bar u,d\bar d,s\bar s$.

Similarly to what has been done in \cite{PR97}, here we
determine photon wave function working in the approach of the
spectral integration technique. This technique had been suggested in
\cite{deut} for the description of deuteron form factors, the deuteron
being treated as a composite two-nucleon system.
In \cite{PR97,PR95},
the spectral integration technique was expanded for
the composite $q\bar q$ systems with wave functions written
in terms of the light-cone variables. The wave function depends on the
invariant energy squared of $q\bar q$ system as follows:
\be
\label{I1}
s=\frac{m^2+k_\perp^2}{x(1-x)}\ ,
\ee
where $m$ is the quark mass and ${\bf k}_\perp$ and $x$  are the
light cone characteristics of quarks (transverse
momentum and a part of longitudinal momentum). In this technique the
quark wave function of the
photon, $\gamma^*(Q^2)\to q\bar q$, is defined as follows:
\be
\label{I2}
\psi_{\gamma^*(Q^2)\to q\bar q}(s)=
\frac{G_{\gamma\to q\bar q}(s)}{s+Q^2}\ ,
\ee
where $G_{\gamma \to q\bar q}(s)$ is the vertex
for the transition of photon into
$q\bar q$ state. Rather schematically the vertex function
$G_{\gamma\to q\bar q}(s)$ may be represented as
\be
\label{I3}
C e^{-bs}+ \theta (s-s_0)\ ,
\label{s0}
\ee
where the first term stands for the soft component  which is due to the
transition of photon to vector $q\bar q$ mesons
$\gamma\to V\to q\bar q$, while the second one describes
the point-like interaction in the hard domain (here
the step-function $\theta(s-s_0)=1$ at $s \geq s_0$ and
$\theta(s-s_0)=0$ at $s<s_0$). The principal characteristics of the
soft component of
$G_{\gamma\to q\bar q}(s)$ is the threshold value of the vertex, $C
\exp(-4m^2b)$, and the rate of its decrease with energy, that is the
slope $b$. The hard component of the vertex is characterised by the
magnitude $s_0$, which is the quark energy squared when the point-like
interaction becomes dominant.

In \cite{PR97}, the photon wave function has been found out assuming
the quark relative-momentum dependence is the same for all quark
vertices: $g_{\gamma\to u\bar u}(k^2)=g_{\gamma\to d\bar d}(k^2)=$
\\$=g_{\gamma\to s\bar
s}(k^2)$, where we redenoted
$G_{\gamma\to q\bar q}(s)\longrightarrow g_{\gamma\to q\bar
q}(k^2)$ with $k^2=s/4-m^2$. The hypothesis of the vertex
universality for $u$ and $d$ quarks,
\be
G_{\gamma\to u\bar u}(s)=G_{\gamma\to d\bar d}(s)\equiv
G_{\gamma}(s)\ ,
\ee
looks rather trustworthy because
of the degeneracy of $\rho$ and $\omega$
states, though the similarity in the $k$-dependence  for the nonstrange
and strange quarks may be violated. In addition, using experimental
data on the transitions $\gamma\gamma^*(Q^2)\to \pi^0,\eta,\eta'$ only
one cannot find out the main parameters  ($C,b,s_0$) both for
$G_{\gamma\to s\bar s}(s)$ and $G_{\gamma}(s)$.
In the present paper we add the $e^+e^-$
annihilation data for the determination of wave functions, that is,
$e^+e^-\to \gamma^*\to \rho^0,\omega,\phi(1020)$ together with the
ratio $R(E_{e^+e^-})=\sigma(e^+e^-\to hadrons)/\sigma(e^+e^-)\to
\mu^+\mu^-)$ at   $E_{e^+e^-}$ higher than 1 GeV. The reactions
$e^+e^-\to \gamma^*\to \rho^0,\omega,\phi(1020)$ are rather sensitive
to the parameters $C,b$ of the soft component of photon wave function,
while the data on $R(E_{e^+e^-})$ allow us to fix
the  parameter $s_0$ for the beginning of the point-like vertex regime,
see Eq. (\ref{s0}).

The paper is organised as follows.
In Section 2, which is in fact the introductory one,
we present the formulae for the charge form
factor of pseudoscalar meson and transition form factors
$\pi^0,\eta,\eta'\to \gamma(Q^2_1)\gamma(Q^2_2) $ in terms of the
spectral integration technigue.
In Section 3 we consider
the $e^+e^-$ annihilation processes: the partial decay widths
$\omega,\rho^0,\phi \to e^+e^-$ and the ratio
$R(E_{e^+e^-})=\sigma(e^+e^-\to
 hadrons)/\sigma(e^+e^-\to \mu^+\mu^-)$ at
$1\le E_{e^+e^-}\le 3.7$ GeV. The photon wave function
$\gamma\to q\bar q$ for the light quarks is determined
in Section 4. The results of calculations for the decays $f_0(980),a_0(980)\to\gamma\gamma$ and
$f_2(1270),f_2(1525),a_2(1320)\to\gamma\gamma$
carried out with
the found here photon wave function are compared with calculations
performed with the old photon wave function \cite{PR97}
in Section 5.  In
Conclusion, we briefly  summarize the results.

\section{Quark--aniquark state form factors in the spectral integration
technique}

In this Section we recall the main formulae for the calculation of
the charge and transition  form
factors in the spectral integration technique, these formulae are
used for the determination of photon wave function.
First, we present
formulae for the charge pion form factor --- they are needed to fix up
the wave function of the pion and other members of
the lowest pseudoscalar nonet, $\eta$ and $\eta'$.
The calculation of the charge form factor is based on a  principal
hypothesis of the additive quark model: the mesons consist of quark and
antiquark, and the photon interacts with one of constituent quarks.
Hereafter
the formulae for the transitions $\pi^0,\eta,\eta'\to
\gamma(Q^2_1)\gamma(Q^2_2)$ are given, they are written   within
similar approach.  More detailed discussion of these formulae and
basic assumptions may be found
in [12-21].

\subsection{Pion charge form factor}

Here  we recall the
logic of calculation in the spectral integration technique
and write down the formulae for pion form factor.

General structure of the amplitude of pion--photon interaction is as
follows:
\be
A^{(\pi)}_\mu=e(p_\mu+p'_\mu) F_\pi(Q^2)\ ,
\label{pi1}
\ee
where $e$ is the absolute value of electron charge, $p$ and $p'$ are
pion incoming--outgoing four-momenta, and $F_\pi(Q^2)$ is the pion form
factor. We are working in the space-like region of the momentum
transfer, so $Q^2=-q^2$, where $q=p-p'$. The amplitude $A^{(\pi)}_\mu$
is the tranverse one: $q_\mu A^{(\pi)}_\mu=0$.

In the quark model, the pion form factor is defined as a process
shown in Fig. 1$a$: the photon interacts with one of constituent quarks.
In the spectral integration technique, the method of calculation of the
diagram of Fig. 1$a$ is as follows: we consider the dispersive
integrals over masses of incoming and outgoing $q\bar q$
states, corresponding cuttings of the triangle diagram are shown in
Fig. 1$b$. In this way we calculate the double discontinuity of the
triangle diagram, ${\rm disc}_s  {\rm disc}_{s'}F_\pi(s,s',Q^2)$, where
$s$ and $s'$ are the energy squares of the $q\bar q$ systems
before and after the photon emission,
$P^2=s$ and $P'^2=s'$ (recall, in the dispersion relation
technique the momenta of intermediate particles
do not coincide with external momenta, $p\ne P$ and $p'\ne P'$).  The
double discontinuity is defined by three factors:\\
(i) product of the
pion vertex functions and quark charge:
\be
e_qG_\pi(s)G_\pi(s') ,
\label{pi2}
\ee
where, due to Eq. (\ref{pi1}), $e_q$ is given
in the units of
the charge $e$,\\
(ii) phase space of the triangle diagram
 (Fig. 1$b$) at $s\geq 4m^2$ and $s'\geq 4m^2$:
\be
d\Phi_{\rm tr}=d\Phi_2(P;k_1,k_2)d\Phi_2(P';k'_1,k'_2)(2\pi)^32k_{20}
\delta^{(3)} (\bk_2-\bk'_2)\ ,
\label{pi3}
\ee
with the two-particle phase space determined as:
\be
d\Phi_2(P;k_1,k_2)=\frac12 \frac{d^3k_1}{(2\pi)^3 2k_{10}}
\frac{d^3k_2}{(2\pi)^3 2k_{20}}(2\pi)^4
\delta^{(4)} (P-k_1-k_2)\ ,
\label{pi3a}
\ee
(iii) spin factor $S_\pi(s,s',Q^2)$ determined by the
trace of the triangle diagram process of Fig. 1$b$:
\be
\label{pi4}
-{\rm Tr}\left [
i\gamma_5(m-\hat k_2)i\gamma_5(m+\hat k_1')\gamma^\perp_\mu(m+\hat k_1)
\right ]=
(P+P')^\perp_\mu S_{\pi}(s,s',Q^2)\ .
\ee
Recall that in the dispersion integral we deal with mass-on-shell
particles, so $k_1^2=k'^2_1=k^2_2=m^2$. The vertex $i\gamma_5 $
corresponds to the transition $\pi\to q\bar q$, the photon is carrying
the momentum $\tilde q=P-P'$,  and the photon momentum square is fixed,
$\tilde q^2=q^2=-Q^2$.
The transversity of the amplitude $A_\mu^{(\pi)}$
is guaranteed by the use of $\gamma_\mu^\perp$ in the trace (\ref{pi4}):
\be
\gamma^\perp_\mu=g^\perp_{\mu\nu} \; \gamma_\nu\, ,
 \label{pi5}
\ee
$$
g^\perp_{\mu\nu} =
g_{\mu\nu}- \frac{(P'_\mu -P_\mu)(P'_\nu -P_\nu)}{q^2 }\ ,
$$
$$
(P+P')^\perp_\mu=
\left[P'_\mu +P_\mu-\frac{P'_\mu -P_\mu}{q^2}(s'-s)\right ]\ .
$$
The spin factor $S_\pi(s,s',Q^2)$ reads:
\be
S_{\pi}(s,s',Q^2)=2\left [(s+s'+Q^2)\alpha(s,s',Q^2)-Q^2\right ]\ ,
 \label{pi6}
\ee
$$
\alpha(s,s',Q^2)=
\frac{s+s'+ Q^2}{2(s+s')+(s'-s)^2/Q^2+Q^2}\ .
$$
As a result, the the double discontinuity of the diagram with a photon
emitted by quark is determined as:
\be
e_q G_\pi(s)G_\pi(s')S_\pi(s,s',Q^2)d\Phi_{\rm tr}\ .
\label{pi8a}
\ee
Emission of the photon by antiquark
gives similar contribution, with a substitution $e_q \to
e_{\bar q}$: so the total charge factor for the $\pi^+$ is the unity,
$e_u + e_{\bar d}=1$. Then the double discontinuity reads:
\be
{\rm disc}_s{\rm disc}_{s'} F_\pi(s,s',Q^2)=
 G_\pi(s)G_\pi(s')S_\pi(s,s',Q^2)d\Phi_{\rm tr}\ .
\label{pi8}
\ee
The form factor $F_\pi(Q^2)$ is defined as a double dispersion integral
as follows:
\be
F_\pi(Q^2)=\inli_{4m^2}^{\infty}\frac{ds}{\pi}\frac{ds'}{\pi}
\, \frac{{\rm disc}_s{\rm disc}_{s'} F_\pi(s,s',Q^2)}
{(s'-m_\pi^2)(s-m_\pi^2)}\ .
\label{pi7}
\ee
When the form factor calculations are performed, it is suitable to
operate with the wave function of composite system. In case of a pion
the wave function is defined as follows:
\be \Psi_\pi(s)=\frac{G_\pi(s)}{s-m_\pi^2} \ .
\label{pi9}
\ee
There are different ways to work with formula (\ref{pi7}),
in accordance to different goals, where the $q\bar q$ system is
involved. Spectral representation of the form factor appears after the
integration in (\ref{pi7}) over the momenta of  constituents by
removing $\delta$-functions in the phase space $d\Phi_{\rm tr}$. Then
\be
F_\pi(Q^2)=\inli_{4m^2}^{\infty}\frac{ds}{\pi}
\frac{ds'}{\pi}\Psi_\pi(s)\Psi_\pi(s')S_\pi(s,s'Q^2)
\frac{\theta \left(s'sQ^2-m^2\lambda(s,s',Q^2)\right)}
{16\sqrt{\lambda(s,s',Q^2)}}\, ,
\label{pi10}
\ee
$$
\lambda(s,s',Q^2)=(s'-s)^2+2Q^2(s'+s)+Q^4\ .
$$
Here the $\theta $-function  determines the integration region over
$s$ and $s'$: $\theta (X)=1$ at $X\geq 0$ and $\theta (X)=0$ at
$X< 0$.

Another way  to present form factor is to remove the integration
over the energy squares  of the quark-antiquark systems,
$s$ and $s'$, by using $\delta$-functions entering $d\Phi_{\rm tr}$.
Then we have
the formula for the pion form factor in the light-cone variables:
\be
F_{\pi}(Q^2 ) = \frac {1}{16\pi^3}
\int \limits_{0}^{1}
\frac {dx}{x(1-x)^2}  \int d^2k_{\perp} \Psi_{\pi} (s)
\Psi_{\pi} (s')
S_{\pi}(s,s',Q^2)\ ,
\label{pi11}
\ee
$$
s=\frac{m^2+k_\perp^2}{x(1-x)}\ , \qquad
s'=\frac{m^2+({\bf k}_\perp-x{\bf Q})^2}{x(1-x)}\ ,
$$
where $\bkp$ and $x$ are the light-cone quark characteristics
(transverse momentum of the quark and a part of momentum along the
$z$-axis).

Fitting formulae  (\ref{pi10})  or (\ref{pi11}) to data
at $0\leq Q^2 \leq 1$ (GeV/{\it c})$^2$ with
two-exponen\-ti\-al parametrization  of the wave function $\Psi_\pi$:
\be
\Psi_\pi(s)=c_\pi\left [\exp(-b^\pi_1 s)+\delta_\pi\exp(-b^\pi_2 s)
\right ]\ ,
\label{pi13}
\ee
we obtain the following values of the pion wave function parameters:
\be
c_\pi=209.36\,{\rm GeV}^{-2} ,\quad \delta_\pi=0.01381,\quad
b^\pi_1=3.57\,{\rm GeV}^{-2}\quad b^\pi_2=0.4\,{\rm GeV}^{-2}\ .
\label{pi14}
\ee
 Figure 2 demonstrates the description  of the data
 by formula (\ref{pi10}) (or (\ref{pi11})) with the pion wave function
 given by
(\ref{pi13}), (\ref{pi14}).

The region $1\leq Q^2 \leq 2$ (GeV/{\it c})$^2$ was not used for the
determination of parameters of the pion wave function: one may
suppose that at $Q^2 \geq 1$ (GeV/{\it c})$^2$  the predictions of additive
quark model fail. However, one can  see that the calculated
curve fits reasonably to data in the neighbouring region $1\leq Q^2
\leq 2$ (GeV/{\it c})$^2$ too (dashed curve in Fig. 2).

The constraint $F_{\pi}(0) = 1$ serves us as a normalization condition
for pion wave function. We have
in the low-$Q^2$ region:
\be
F_{\pi}(Q^2 ) \simeq 1-\frac{1}{6}R_{\pi}^2 Q^2\ ,
\label{pi12}
\ee
with  $R_{\pi}^2\simeq 10$  (GeV/{\it c})$^{-2}$.
The pion radius is just the characteristics, which is used later on for
comparative estimates  of the wave function parameters for
other low-lying $q\bar q$ states.

\subsection{Transition form factors  $\pi^0, \eta,\eta' \to \gamma
^*(Q_1^2 ) \gamma ^*(Q_2^2 )$.}

Using the same technique we can write the formulae for transition form
factors of pseudoscalar mesons
$\pi^0, \eta,\eta' \to \gamma
^*(Q_1^2 ) \gamma ^*(Q_2^2 )$, the corresponding diagrams are shown\\ in
Fig. 3.

For these processes, general structure of
the amplitude  is as follows:
\be
\label{31}
A_{\mu\nu}(Q_1^2,Q_2^2)=e^2\epsilon_{\mu\nu\al\bet}q_\al
p_\bet F_{(\pi,\eta,\eta')\to \gamma \gamma }(Q_1^2,Q_2^2)\ .
\ee
In the light-cone  variables $(x,\bkp)$,
the expression for the transition form factor
$\pi^0\to \gamma ^*(Q^2_1)\gamma ^*(Q^2_2) $ determined
by two processes of Fig. 3$a$ and Fig. 3$b$  reads:
\be
\label{32}
F_{\pi\to \gamma \gamma }(Q_1^2,Q_2^2)=
\zeta_{\pi\to \gamma \gamma }
\frac{\sqrt{N_c}}{16\pi^3} \inli_0^1
\frac{dx}{x(1-x)^2}\int d^2k_\perp \Psi_{\pi}(s)\times
\ee
$$\times \left(S_{\pi\to \gamma \gamma }(s,s'_1,Q^2_1)
 \frac{G_\gamma
(s'_1)}{s'_1+Q^2_2}+
S_{\pi\to \gamma \gamma }
(s,s'_2,Q^2_2)\frac{G_\gamma (s'_2)}{s'_2+Q^2_1}
\right )\ ,
$$
where
\be
s=\frac{m^2+k_\perp^2}{x(1-x)}\ , \qquad
s'_i=\frac{m^2+({\bf k}_\perp-x{\bf Q}_i)^2}{x(1-x)}\ ,\qquad (i=1,2).
\ee
 The spin factor for pseudoscalar states depends on the quark mass
only:
\be
\label{33}
S_{\pi\to \gamma \gamma }(s,s'_i,Q^2) =4m \, .
\ee
The charge factor for the decay $\pi^0\to \gamma \gamma  $ is equal to
\be
\label{34}
\zeta_{\pi\to \gamma \gamma }=\frac{e^2_u-e^2_d}{\sqrt 2}=
\frac{1}{3\sqrt 2} \, .
\ee
In (\ref{32}) the ratio
$G_\gamma (s_i)/(s_i+Q^2)$ is the photon wave function (remind,
we denote
$G_{\gamma\to u\bar u}(s)=G_{\gamma\to d\bar d}(s)\equiv G_\gamma (s)$).
The factor $\sqrt{N_c}$ in the right-hand side of (\ref{32}) appears
due to another definition of the colour wave function of the
photon as compared to pion's one: without $1/\sqrt{N_c}$.

In terms of the spectral
integrals over the $(s,s')$ variables, the transition form factor
for $\pi^0\to \gamma^*(Q_1^2) \gamma^*(Q_2^2)$ reads:
\be
\label{35}
 F_{\pi\to \gamma \gamma } (Q_1^2,Q_2^2)= \zeta_{\pi\to \gamma \gamma
}\frac{\sqrt{N_c}}{16}\int \limits_{4m^2}^\infty
\frac{ds}{\pi}\frac{ds'}{\pi} \Psi_\pi(s)  \times
\ee
$$
\times\left
[\frac{\theta(s'sQ_1^2-m^2\lambda(s,s',Q_1^2))}{\sqrt{\lambda
(s,s',Q_1^2)}}S_{\pi}(s,s',Q^2_1) \frac{G_\gamma (s')}{s'+Q^2_2}+
\right.$$
$$\left.+
\frac{\theta(s'sQ_2^2-m^2\lambda(s,s',Q_2^2))}{\sqrt{\lambda
(s,s',Q_2^2)}}S_{\pi}(s,s',Q^2_2) \frac{G_\gamma (s')}{s'+Q^2_1}
\right ]\ ,
$$
where $\lambda(s,s',Q_i^2)$ is determined in (\ref{pi10}).

Similar expressions may be written for the transitions
$\eta,\eta'\to \gamma^*(Q_1^2) \gamma^*(Q_2^2)$.
One should bear in mind that, because of the presence of two
quarkonium components, their flavour wave functions are:
$s\bar s$ and
 $n\bar n=(u\bar u+d\bar
d)/\sqrt{2}$, in the $\eta$, $\eta'$-mesons,
$$
\eta=\sin \theta\, n\bar n-\cos \theta\, s\bar s, \quad
 \eta'=\cos \theta\, n\bar n+\sin \theta\, s\bar s,
$$
their transition form factors are
expressed through mixing angle $\theta$ as follows:
\bea
F_{\eta\to \gamma \gamma }(s)&=&
\sin \theta F_{\eta /\eta' (n\bar n)\to \gamma \gamma}(s)-
\cos \theta
F_{\eta /\eta' (s\bar s)\to \gamma \gamma}(s)\ ,
\\ \nonumber
 F_{\eta'\to \gamma \gamma }(s)&=&
\cos \theta F_{\eta /\eta'(n\bar n)\to \gamma \gamma}(s)+
\sin \theta F_{\eta /\eta'(s\bar s)\to \gamma \gamma }(s)\ .
\label{36}
\eea
The spin factors  for nonstrange
components of $\eta$ and $\eta'$ are the same as those for the pion,
$S_{\eta /\eta' (n\bar n)\to \gamma \gamma}(s,s',Q^2)=4m$,
though with another quark mass  entering  the
strange com\-po\-nent:
 and
$S_{\eta /\eta' (s\bar s)\to \gamma \gamma}(s,s',Q^2)=4m_s$.

 Charge factors for the $n\bar n$ and $s\bar s$ components are equal to
\be
\label{37}
\zeta_{\eta/\eta'(n\bar n)\to \gamma\gamma}=\frac {5}{9\sqrt{2}}, \qquad
\zeta_{\eta/\eta'(s\bar s)\to \gamma\gamma}=\frac 19.
\ee
In the calculation of transition form factors of pseudoscalar mesons,
the wave function related to nonstrange quarks  in
$\eta$ and $\eta'$ was
assumed to be the same as for the pion:
 $$\Psi_{\eta/\eta'(n\bar n)}(s)=\Psi_{\pi}(s)\ .
\eqno{(28a)}
$$
 As to  strange components
of the wave functions,
we suppose the same shape for
$n\bar n$ and $s\bar s$. For
 $\Psi_{\eta/\eta'(s\bar s)}(s)=\Psi_{\pi}(s)$,
it results
in another  normalization only as compared to (28a):
\be
\Psi_{\eta/\eta'(s\bar s)}(s)=c_{\eta/\eta'(s\bar s)}
\left [\exp(-b^{\eta/\eta'(s\bar s)}_1 s)+
\delta_{\eta/\eta'(s\bar s)}\exp(-b^{\eta/\eta'(s\bar s)}_2 s)
\right ]\ ,
\label{38}
\ee
$$
c_{\eta/\eta'(s\bar s)}=528.78\, \ {\rm GeV}^{3/2}\, ,\quad
\delta_{\eta/\eta'(s\bar s)}=\delta_\pi\, ,\quad
b^{\eta/\eta'(s\bar s)}_1=b^\pi_1\, ,\quad
b^{\eta/\eta'(s\bar s)}_2=b^\pi_2 \ .
$$

\section{$e^+e^-$-annihilation}

The $e^+e^-$-annihilation processes provide us
with additional information on the photon wave function:

\noindent
(i) Partial width of the transitions $\omega,\rho^0,\phi\to e^+e^-$ is
defined  by the quark loop diagrams, which contain the product
$G_\gamma(s)\Psi_V(s)$, where $\Psi_V(s)$ is the quark wave function of
vector meson ($V=\omega,\rho^0,\phi$). Supposing that  radial wave
functions of  $\omega,\rho^0,\phi$  coincide with
those of the
lowest pseudoscalar mesons (this assumption looks verisimilar, for
these mesons
are  members of the same lowest 36-plet), we can obtain information
about $G_\gamma(s)$ and $G_{\gamma(s\bar s)}(s)$ from the data on the
$\omega,\rho^0,\phi\to e^+e^-$ decays \cite{PDG}.

\noindent
(ii) The ratio
$R(s)=\sigma(e^+e^-\to hadrons)/\sigma(e^+e^-\to \mu^+\mu^-)$ at
high center-of-mass energies but below the open charm production
($\sqrt{s}\equiv E_{e^+e^-}<3.7$ GeV) is
determined by hard components of the photon vertices $G_\gamma(s)$ and
$G_{\gamma(s\bar s)}(s)$ (transitions $\gamma^*\to u\bar u,d\bar
d,s\bar s$), thus giving us  a well-known magnitude $R(s)=2$ (small
deviations from $R(s)=2$ comes from  corrections related  to the
gluon emission  $\gamma^*\to q\bar q g$,  see \cite{ryskin}
and references therein). Hence the deviation of
the ratio from the value $R(s)=2$ at decreasing
$E_{e^+e^-}$ provides us with the information about the energies, when
the regime changes: hard components in $G_\gamma(s)$ and
$G_{\gamma(s\bar s)}(s)$ stop working, while soft components start to
play their role.

\subsection{ Partial decay widths $\omega,\rho^0,\phi\to e^+e^-$}

Figure 4 is a diagrammatic representation of the reaction $V\to
e^+e^-$: virtual photon produces the $q\bar q$ pair, which  turns into
 vector meson.

Partial width  of vector meson is determined as follows:
\be
\label{41}
m_V\Gamma_{V\to e^+e^-}
= \pi \alpha^2\, A^2_{e^+ e^-\to V} \,\frac1{m_V^4}
\left(\frac43m_V^2+\frac83m^2_e\right)
\sqrt{\frac{m^2_V-4m_e^2}{m_V^2}}\ .
\ee
Here  $m_V$ is the vector meson mass, the factor $1/m^2_V$ is
associated with photon propagator,
and $\alpha=e^2/(4\pi)$.  In (\ref{41}),
the integration over electron-positron phase space results in
$\sqrt{(m^2_V-4m_e^2)/m_V^2}/(16\pi)$, while the averaging over vector
meson polarizations  and summing over electron--positron  spins  gives:
\be
\label{42}
\frac13\, {\rm Tr}\left[\gamma_\mu^\perp(\hat
k_1+m_e) \gamma_\mu'^\perp(-\hat k_2+m_e)\right]  =
\frac43m_V^2+\frac83m^2_e\ .
\ee
The amplitude
$A_{V\to e^+ e^-}$ is determined through
 the quark--antiquark loop
calculations, within spectral-integration technique.
In this way, we get for the decays  $\omega,\rho^0\to e^+e^-$:
\be
\label{42a}
A_{\omega ,\rho^0\to e^+e^-}=Z_{\omega ,\rho^0}
\frac{\sqrt{N_c}}{16\pi}\int\limits_{4m^2}^{\infty}
\frac {ds}{\pi} G_\gamma(s)\Psi_{\omega ,\rho}(s)
\sqrt{\frac{s-4m^2}{s}}\left(\frac83 m^2+\frac43 s\right )\ ,
\ee
where $Z_{\omega ,\rho^0}$ is the  quark charge factor for vector
mesons:
 $Z_\omega= 1/(3\sqrt{2})$ and $Z_{\rho^0}  =1/\sqrt{2}$.
We have similar expression for the $\phi (1020)\to e^+e^-$ amplitude:
\be
\label{43}
A_{\phi \to e^+e^-}=Z_{\phi } \frac{\sqrt{N_c}}{16\pi}
\int\limits_{4m^2_s}^{\infty}
\frac {ds}{\pi} G_{\gamma\to s\bar s}(s)\Psi_{\phi }(s)
\sqrt{\frac{s-4m^2_s}{s}}\left(\frac83 m^2_s+\frac43 s\right )\ ,
\ee
with $Z_\phi  = 1/3$.
The normalization condition for the vector-meson wave function reads:
\be
\label{45}
\frac1{16\pi}
\int\limits_{4m^2}^{\infty}
\frac {ds}{\pi} \Psi_{V}^2(s)
\sqrt{\frac{s-4m^2}{s}}\left(\frac83 m^2+\frac43 s\right )\ =1\ .
\ee
The wave function is parametrized in one-exponential form:
\be
\Psi_{V}(s)=c_{V} \exp(-b_{V} s) \label{44} \ ,
\ee
with
\be
b_{\omega,\rho}=2.2\,{\rm GeV}^{-2} ,\quad  c_{\omega ,\rho}=95.1
 \, {\rm GeV}^{-2}
\label{44a}
\ee
for the non-strange mesons and
\be
b_{\phi}=2.5\, {\rm GeV}^{-2} , \quad c_{\phi}(s\bar s)=374.8  \,
{\rm GeV}^{-2}
\ee
for the $\phi (1020)$. Within the used parametrization
the vector mesons are characterized by the following
mean radii squared:  $R_{\omega,\rho}^2=10\, {\rm (GeV/{\it c})}^{-2}$
and $ R_{\phi}^2=11\, {\rm (GeV/{\it c})}^{-2}$.

\subsection{The ratio $R(s)=\sigma(e^+e^-\to
 hadrons)/\sigma(e^+e^-\to \mu^+\mu^-)$ at energies below the
 open charm production}

 At high energies but below the open charm production,
 $E_{e^+e^-}=\sqrt{s}<3.7$ GeV, the ratio $R(s)$ is determined by the
 sum of quark charges squared
in the transition
 $e^+e^-\to\gamma^*\to u\bar u+d\bar d+s\bar s$ multiplied
 by the factor $N_c=3$:
 \be
R(s)=\frac{\sigma(e^+e^- \to
hadrons)}{\sigma(e^+e^-\to\mu^+\mu^-)}=N_c(e_u^2+e_d^2+e_s^2)=2\ .
 \label{47}
 \ee
Since  the $G_\gamma(s)$ and $G_{\gamma(s\bar s)}(s)$
vertices are  normalized  as $G_\gamma(s)=G_{\gamma(s\bar s)}(s)=1$ at
 $s\to \infty$, we can relate   at large $s$ $R(s)$ and
 \be
 R_{\rm vert}(s) =
3(e_u^2+e_d^2)G_\gamma^2(s)+3e_s^2G_{\gamma(s\bar s)}^2(s)
 = \frac53 G_\gamma^2(s)+\frac13
 G_{\gamma(s\bar s)}^2(s)
 \label{48}
 \ee
  to each other:
 \be
 \label{49}
 R(s)\simeq R_{\rm vert}(s)\ .
 \ee
 Following (\ref{49}), we determine
 the energy region, where the hard components in
 $G_\gamma(s)$, $G_{\gamma(s\bar s)}(s) $ start to dominate.

\section{ Photon wave function }

To determine the photon wave function
we use:
\noindent
(i) transition widths
 $\pi^0, \eta,\eta'\to \gamma\gamma^*(Q^2)$,
\noindent
(ii) partial decay widths
 $\omega,\rho^0,\phi\to e^+e^- ,\, \mu^+\mu^-$,
\noindent
(iii) the ratio
$R(s)=\sigma(e^+e^-\to hadrons)/\sigma(e^+e^-\to \mu^+\mu^-)$.

Transition vertices for $u\bar u,d\bar d\to\gamma$ and
$s\bar s\to\gamma$ have
been chosen in the following form:
\bea
u\bar u,\, d\bar d : \qquad
G_\gamma(s)&=&c_\gamma\left (e^{-b^\gamma_1 s}+c^\gamma_2
e^{-b^\gamma_2 s}\right )+\frac1{1+e^{-b^\gamma_0(s-s_0^\gamma)}}\ ,
\nonumber \\
s\bar s : \qquad
G_{\gamma (s\bar s)}(s)&=&c_{\gamma (s\bar s)}
 e^{-b^{\gamma(s\bar s)}_1 s}
+\frac1{1+e^{-b^{\gamma(s\bar s)}_0(s-s_{0}^{\gamma (s\bar s)})}}\ .
\label{51}
\eea
Recall that  photon wave function is determined as
$\Psi_\gamma(s,Q^2)=G_\gamma(s)/(s+Q^2)$, see Eq. (\ref{I2}).

The following parameter values  have been found in fitting to data:
\bea
\label{52}
u\bar u,\, d\bar d: \,\,         && c^\gamma=32.577,\;
c^\gamma_2=-0.0187,\;b^\gamma_1=4\, {\rm GeV}^{-2},\;
b^\gamma_2=0.8\, {\rm GeV}^{-2},\;   \\
&& b^\gamma_0=15\,{\rm GeV}^{-2},\; s_0^\gamma=1.62\,{\rm GeV}^{2}\ ,
\nonumber \\
s\bar s: \,\,
         &&  c_{\gamma( s\bar s)}=310.55,\;
b^{\gamma( s\bar s)}_1=4\,{\rm GeV}^{-2},\;
b^{\gamma (s\bar s)}_0=15\,{\rm GeV}^{-2},\;
s_{0}^{\gamma (s\bar s)}=2.15\,{\rm GeV}^{2}.
\nonumber
\eea

Now let us present the results of the fit in a more detail.

Figure 5 shows the data for
 $\pi^0\to \gamma\gamma^*(Q^2)$ \cite{CELLO,PDG},
$ \eta\to \gamma\gamma^*(Q^2)$ [4-6,22]
and $\eta'\to \gamma\gamma^*(Q^2)$ [2,4-6,22]
We perform the  fitting procedure in the interval $0\le Q^2\le 1$
(GeV/{\it c})$^2$, the fitting curves are shown by solid lines.  The
continuation of the  curves into  neighbouring region
 $1\le Q^2\le 2$ (GeV/{\it c})$^2$
(dashed lines) demonstrates that there is also
reasonable description of the data.

The calculation results for the $V\to e^+e^-$ decay partial widths
{\it versus } data \cite{PDG} are shown below (in keV):
\bea
\Gamma^{\rm calc}_{\rho^0\to e^+e^-}=7.50\;\ ,   \qquad
\Gamma^{\rm exp}_{ \rho^0\to e^+e^-}= 6.77\pm 0.32\;  \ ,
\nonumber \\
\Gamma^{\rm calc}_{ \omega\to e^+e^-}=0.796\;\ ,   \qquad
\Gamma^{\rm exp}_{ \omega\to e^+e^-}=0.60\pm 0.02\;   \ ,
\nonumber \\
\Gamma^{\rm calc}_{ \phi\to e^+e^-}=1.33\;\ ,   \qquad
\Gamma^{\rm exp}_{ \phi\to e^+e^-}= 1.32\pm 0.06\;  \ ,
\nonumber \\
\Gamma^{\rm calc}_{ \rho^0\to \mu^+\mu^-}=7.48\;\ ,   \qquad
\Gamma^{\rm exp}_{ \rho^0\to \mu^+\mu^-}=6.91\pm 0.42\;   \ ,
\nonumber \\
\Gamma^{\rm calc}_{ \phi\to \mu^+\mu^-}=1.33\;\ ,   \qquad
\Gamma^{\rm exp}_{ \phi\to \mu^+\mu^-}= 1.65\pm 0.22\;  \ .
\label{53}
\eea

Figure 6$a$ demonstrates the data for $R(s)$ \cite{ryskin}
 at $E_{e^+e^-} >1$  GeV (dashed area)  versus
 $R_{\rm vert}(s)$ given by Eq. (\ref{48})
with parameters (\ref{52}) (solid line).

In Fig. 6$b$,$c$ one can see the $k^2$-dependence of
 photon wave functions  on the
quark relative momentum square
$k^2$ (here $s=4m^2+4k^2$) for the nonstrange  and strange
components found in our fit (solid line) and that found
in [9] (dashed lines).
One may see that in the region $0\le k^2\le 2.0$ (GeV/{\it c})$^2$ the
scrupulous distinction is rather considerable, though in the
average the old and new wave functions almost coincide. In the next
section we compare the results obtained for the two-photon decays of
scalar and tensor mesons , $S\to \gamma\gamma$ and $T\to \gamma\gamma$
calculated with old and new wave functions.

\section{Transitions  $S\to \gamma\gamma$ and $T\to \gamma\gamma$ }

As was mentioned above, in the average the old  \cite{PR97} and new
photon wave functions coincide, though they differ in  details. So
it would be useful to understand  to what extent this difference
influences the calculation results for the two-photon decays of scalar
and tensor mesons.

The calculation of the two-photon decays of scalar mesons
$f_0(980)\to \gamma\gamma$ and $a_0(980)\to \gamma\gamma$ have been
performed in \cite{epja,YFscalar} with old wave function, under the
assumption that $f_0(980)$ and $a_0(980)$ are the $q\bar q$ systems.
The results of the calculations are shown in Fig. 7 (dashed line). Solid
curve shows the values found with new photon  wave function; for
$a_0(980)$, new wave function reveals stronger dependence on the radius
squared as compared to the old wave function. In the region
$R^2_{a_0(980)}\sim R^2_\pi=10$ (GeV/{\it c})$^{-2}$, the value $\Gamma\left
(a_0(980)\to \gamma\gamma\right )$ calculated with new wave function
becomes 1.5--2 times smaller as compared to old wave function --- this
reflects more precise definition of photon wave function. In the
calculations, the flavour wave function of $f_0(980)$ was defined as
folows:
 $$f_0(980):\qquad n\bar n\,\cos\varphi+s\bar s\,\sin\varphi\ .
$$
In Fig. 8, the calculated areas are shown for the region $\varphi<0$
that is governed by the $K$-matrix analysis results of meson spectra
\cite{kmat,ufn02}. Shaded area  responds to the experimental data
\cite{pennington}.

The $f_0(980)$ being the $q\bar q$ system is characterized by two
parameters: the mean radius squared of $f_0(980)$ and mixing angle
$\varphi$. In Fig. 9 the areas allowed for these parameters are shown;
they were obtained for the processes  $f_0(980)\to \gamma\gamma$ and
$\phi(1020)\to \gamma f_0(980)$ with old photon wave function (Fig.
 9$a$) and new one (Fig. 9$b$). The change of the allowed areas
$(R^2_{f_0(980)},\varphi)$ for the reaction  $f_0(980)\to \gamma\gamma$
is rather noticeable, but it should be emphasized that it does not lead
to a cardinal alteration of  the parameter magnitudes.

Another set of reactions calculated with photon wave function is the
two-photon decay of tensor mesons as follows
$a_2(1320)\to \gamma\gamma$, $f_2(1270)\to \gamma\gamma$    and
$f_2(1525)\to \gamma\gamma$. The calculations of
$a_2(1320)\to \gamma\gamma$ with old and new wave functions are shown
in Fig. 10 (dotted and solid lines, respectfully), experimental data
\cite{PDG,pennington} are presented in Fig. 10 too (shaded areas). The
description of experimental data  has been carried out at
$R^2_{a_2(1320)}\sim 8$ (GeV/{\it c})$^{-2}$: in this region the
 difference
between the calculated values of partial widths, which is due to a
change of wave function, is of the order of 10--20\%.

The amplitude of the transition $f_2\to \gamma\gamma$ is determined by
four form factors related to existence of
two flavour components and two spin structures,
see \cite{epja,YFtensor} for the details.
The calculations of these four form factors
with old and new wave functions are shown in Fig. 11 --- at
$R^2_T\sim $8--10 (GeV/{\it c})$^{-2}$ the difference is of the order of
of 10--20\%.
In Fig. 12, we
show the allowed areas $(R^2_T,\varphi_T)$ obtained in the
description of experimental widths
$\Gamma \left( f_2(1270)\to \gamma\gamma \right)$  and
$\Gamma \left( f_2(1525)\to \gamma\gamma \right)$  \cite{PDG} with
old (Fig. 12$a$) and new (Fig. 12$b$) wave functions. The new photon
wave function results in a more rigid constraint for the areas
$(R^2_T,\varphi_T)$, though there is no qualitative changes in the
description of data. The two-photon
decay data give us two solutions for the
$(R^2_T,\varphi_T)$-parameters:
\begin{eqnarray}
 (R^2_T,\varphi_T)_I&\simeq&\left(8\,{\rm (GeV/{\it c})}^{-2},0\right )\ ,\\
\nonumber
 (R^2_T,\varphi_T)_{II}&\simeq&\left(8\,{\rm
 (GeV/{\it c})}^{-2},25^\circ\right ) \ .
\end{eqnarray}
 The solution with $\varphi\simeq 0$, when $f_2(1270)$ is nearly pure
 $n\bar n$ state and $f_2(1525)$   is an $s\bar s$ system, is more
 preferable from the point of view of hadronic decays as well as the
 analysis \cite{L3sar}.

\section{Conclusion}

 Meson--photon transition form factors have been discussed within
 various approaches such as perturbative QCD formalism \cite{Brodsky,
 Cao}, QCD sum rules [30-32],
variants of the light-cone quark model [9,33-37].
 A distinctive feature of the quark model approach \cite{PR97} consists
 in taking account of soft interaction of quarks in the subprocess
 $\gamma\to q\bar q$, that is, the acoount for the production of vector
 mesons in the intermediate state: $\gamma\to V\to q\bar q$.

 In the present paper
we have re-analysed the quark components of the photon wave function
(the $\gamma^*(Q^2)\to u\bar u, d\bar d, s\bar s$ transitions) on the
basis  of data on the reactions
$\pi^0,\eta,\eta'\to\gamma\gamma^*(Q^2)$, $e^+e^-\to\rho^0,\omega,\phi$
and $e^+e^-\to hadrons$. On the qualitative level, the wave function
obtained here coincide with  that defined before \cite{PR97} by using
the transitions $\pi^0,\eta,\eta'\to\gamma\gamma^*(Q^2)$ only. The data
on the reactions $e^+e^-\to\rho^0,\omega,\phi$ and $e^+e^-\to hadrons$
allowed us to get the wave function structure more precise, in
particular, in the region of the relative quark momenta $k\sim 0.4-1.0$
GeV/{\it c}.

Such an improvement of our knowledge of the photon wave function does
not lead to a cardinal change in the description of two-photon
decays, basis scalar and tensor mesons obtained before [12-14].
Still, a more detailed definition of
photon wave function is important for the calculations of the decays of
a loosely bound $q\bar q$ state such as radial excitation state
or reactions with virtual photons, $q\bar q\to
\gamma^*(Q^2_1)\gamma^*(Q^2_2)$.

We thank M.G. Ryskin for useful discussions.

This work was supported by the Russian Foundation for Basic Research,
project no. 04-02-17091.

\newpage

\begin{figure}[h]
\centerline{\epsfig{file=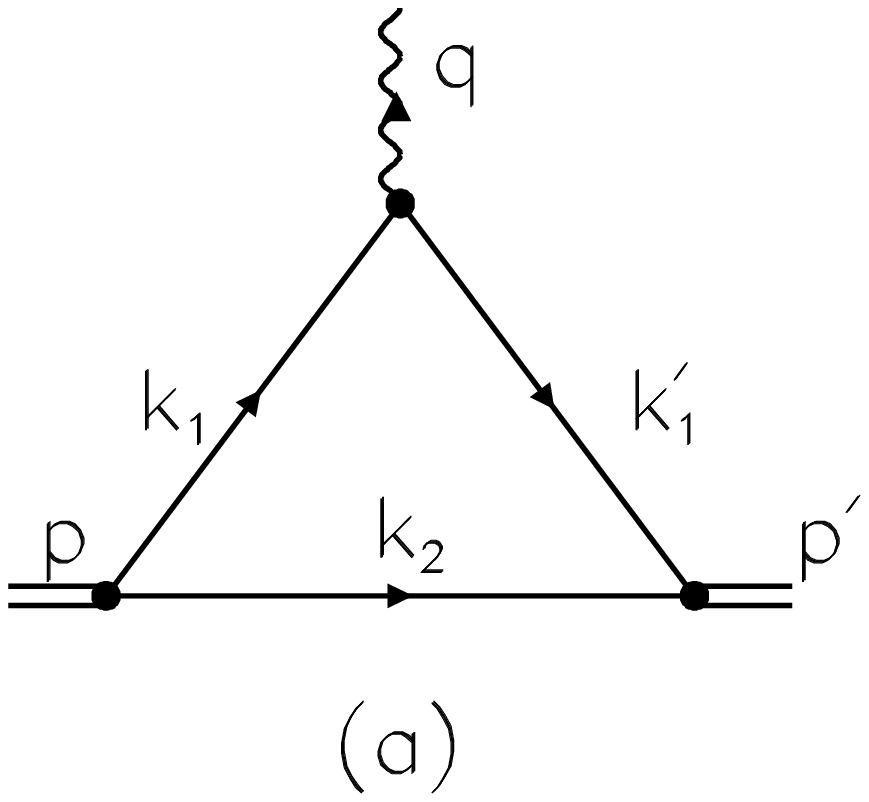,width=6cm}\hspace{0.5cm}
            \epsfig{file=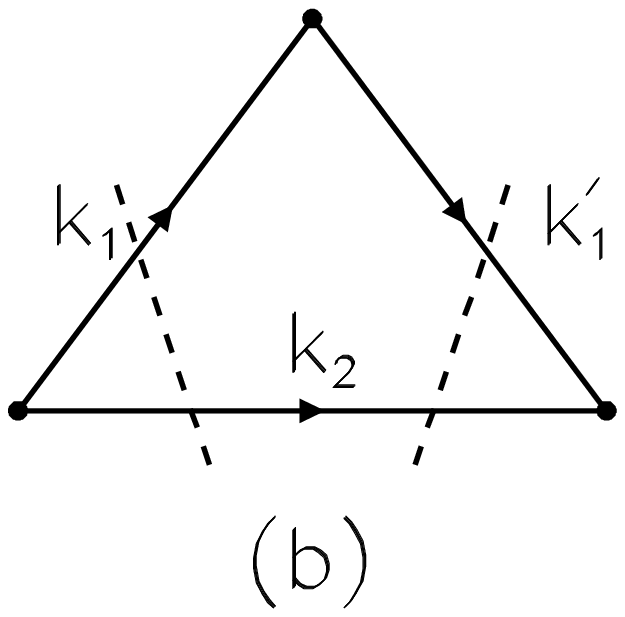,width=6cm}}
\caption{($a$) Diagram for the meson charge form factor in the
additive quark model.  ($b$) Cuts of  triangle diagram in the
spectral integral representation.} \vspace{1cm} \end{figure}

\begin{figure}[h]
\vspace{-1cm}
\centerline{\epsfig{file=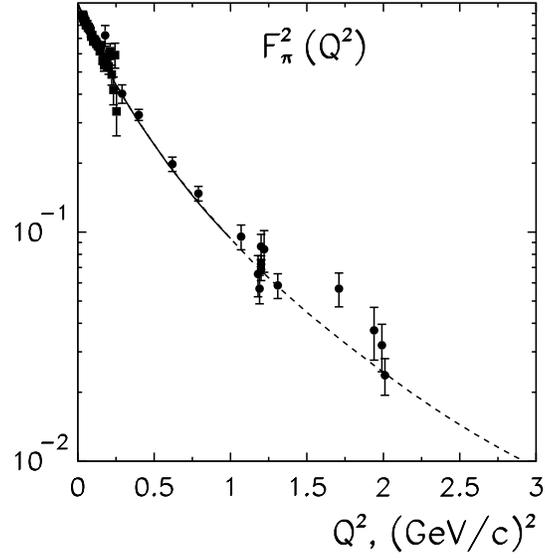,width=8cm}}
\caption{Description of the experimental data on pion charge form
factor with pion wave function given by (18), (\ref{pi14}).}
\end{figure}

\newpage

\begin{figure}[h]
\centerline{\epsfig{file=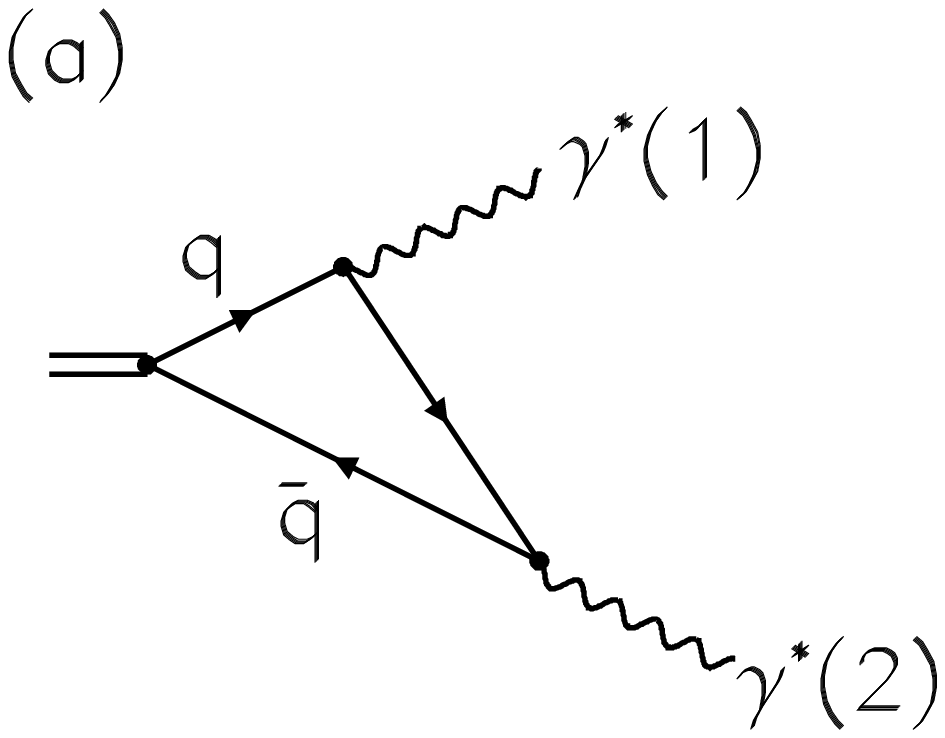,width=6cm}\hspace{0.5cm}
            \epsfig{file=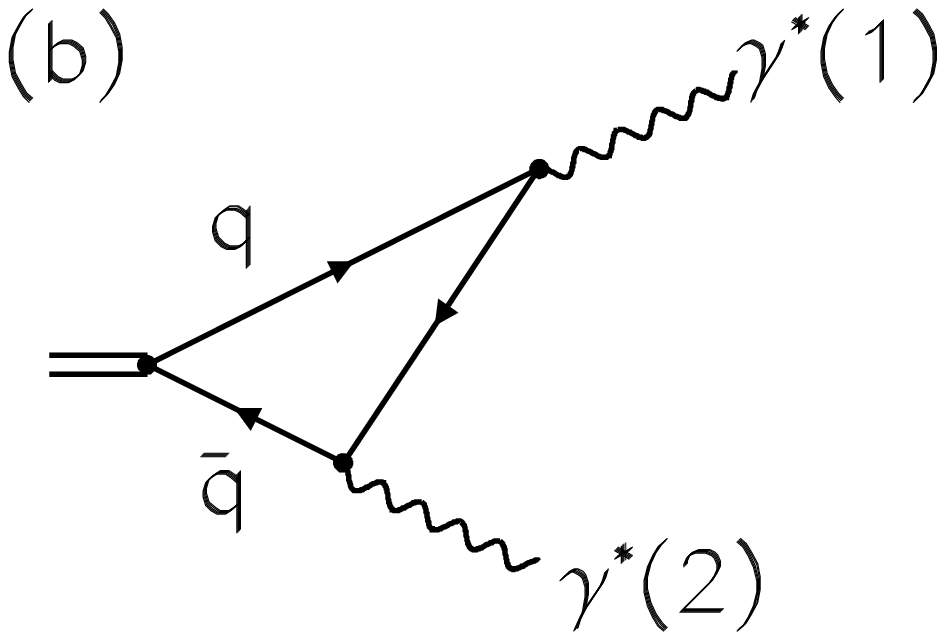,width=6cm}}
\caption{Diagrams for the two-photon decay of $q\bar q$ state with
the emission of photon in the intermediate state by quark ($a$) and
antiquark ($b$). }
\end{figure}

\begin{figure}
\centerline{\epsfig{file=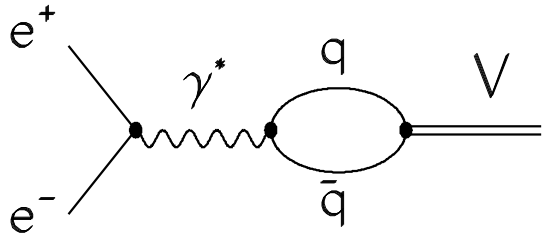,width=9cm}}
\caption{Production of vector $q\bar q$ state in the
$e^+e^-$-annihilation. $V$ denotes the vector meson.}
\end{figure}

\newpage

\begin{figure}
\centerline{\epsfig{file=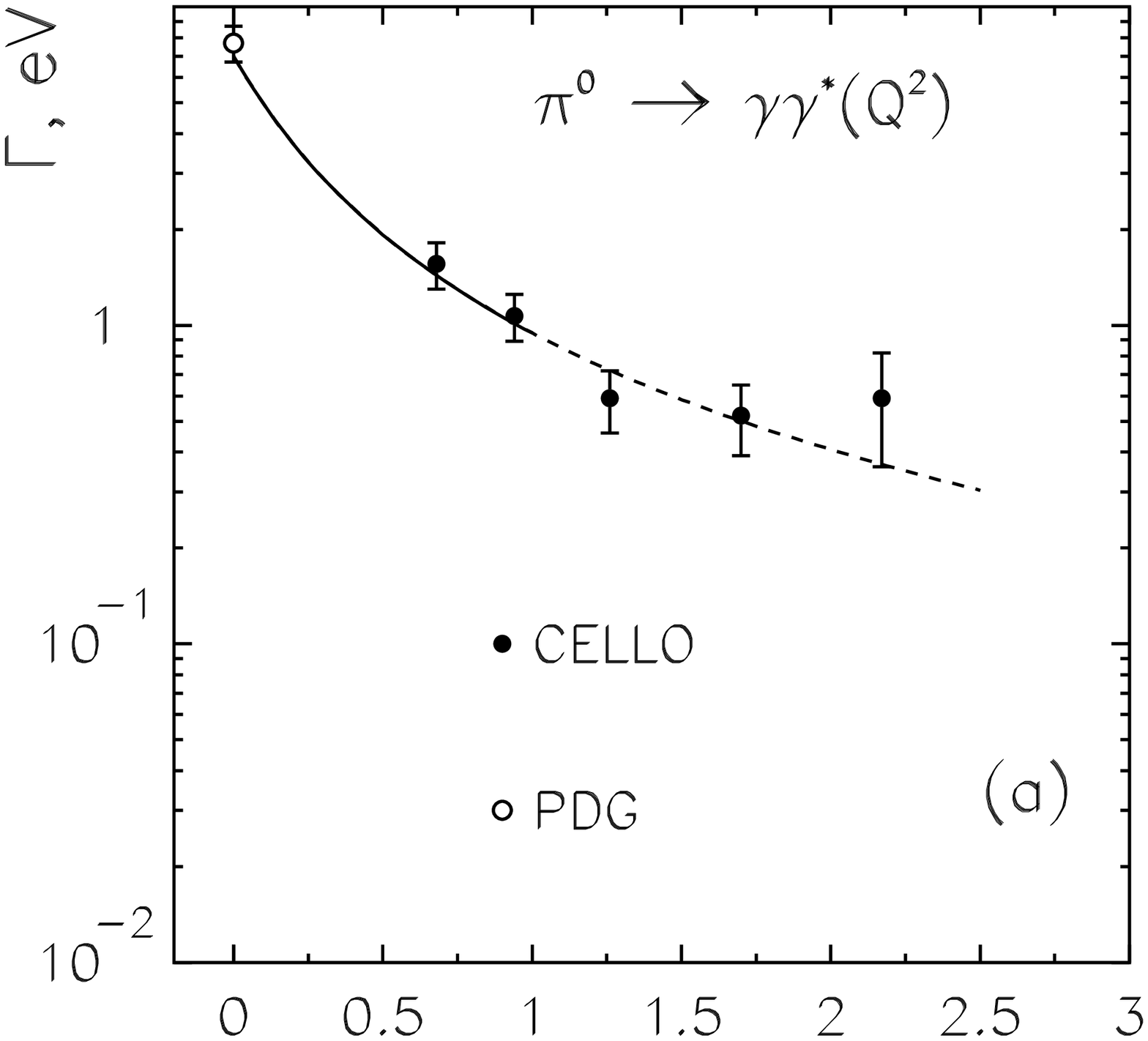,width=7cm}}
\vspace{-1.0cm}
\centerline{\epsfig{file=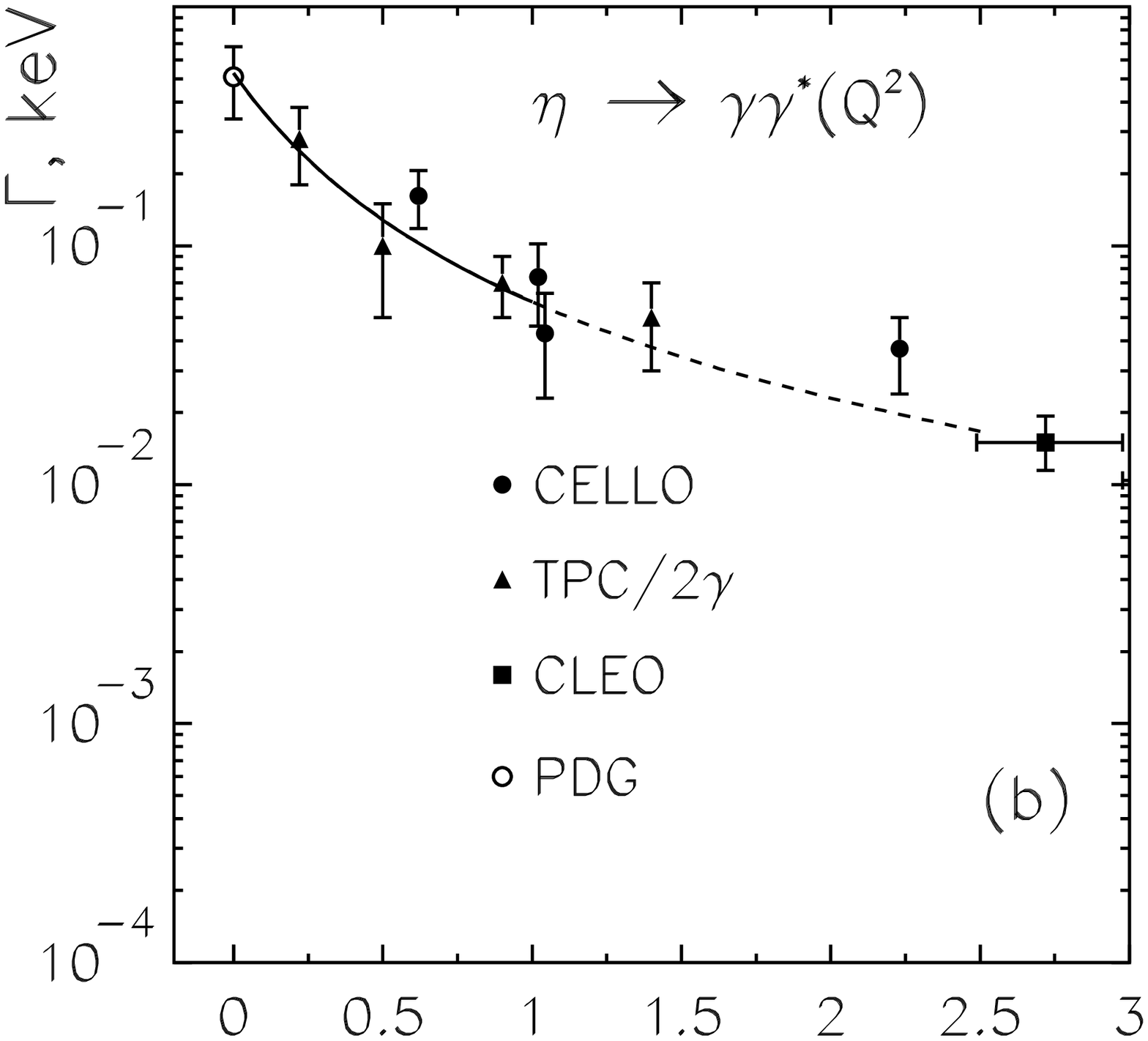,width=7cm}}
\vspace{-1.0cm}
\centerline{\epsfig{file=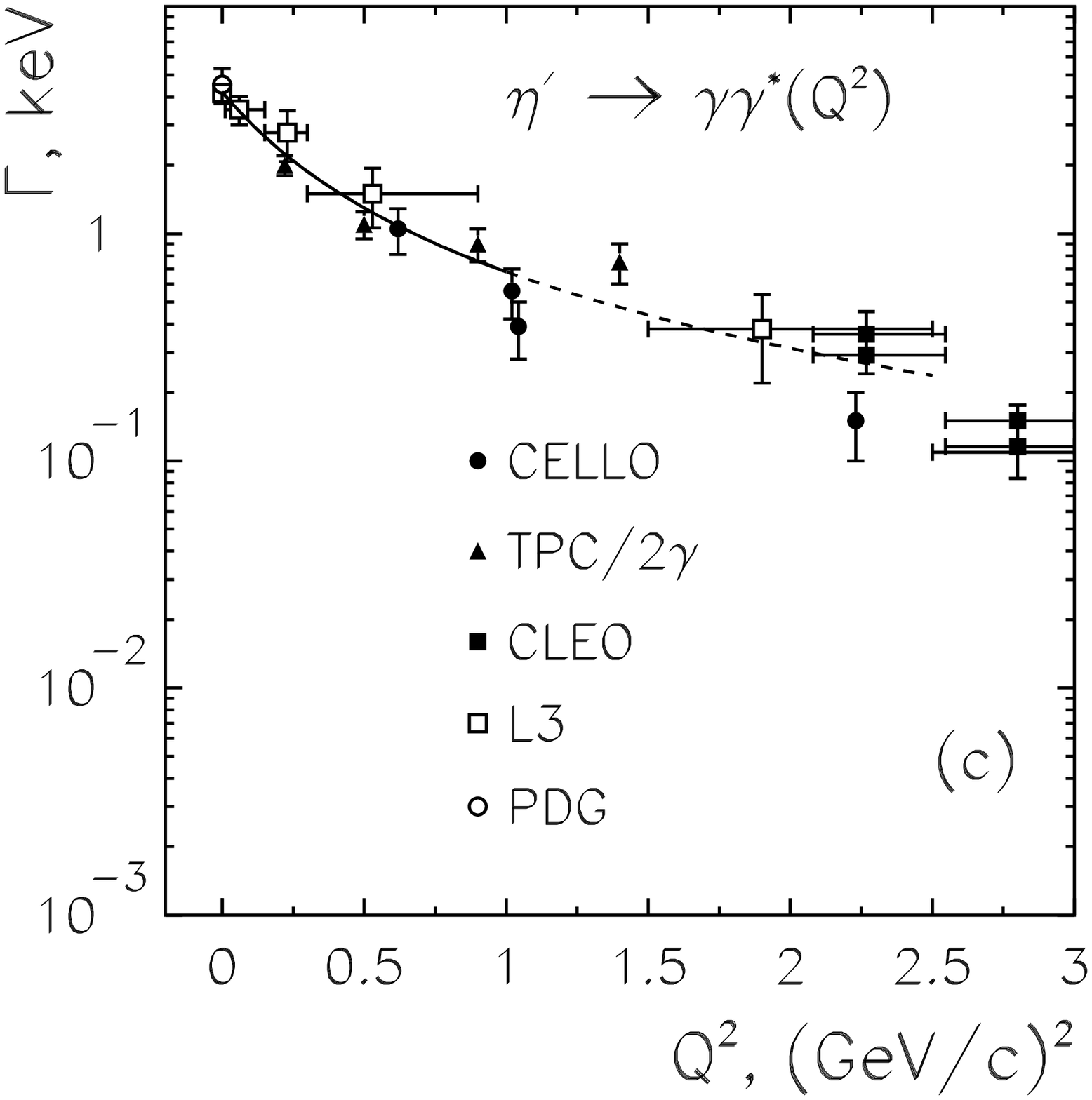,width=7cm}}
\caption{Data for $\pi^0\to \gamma\gamma^*$,
$\eta\to \gamma\gamma^*$ and $\eta'\to \gamma\gamma^*$ {\it vs}
the calculation curves.}
\end{figure}

\begin{figure}
\centerline{\epsfig{file=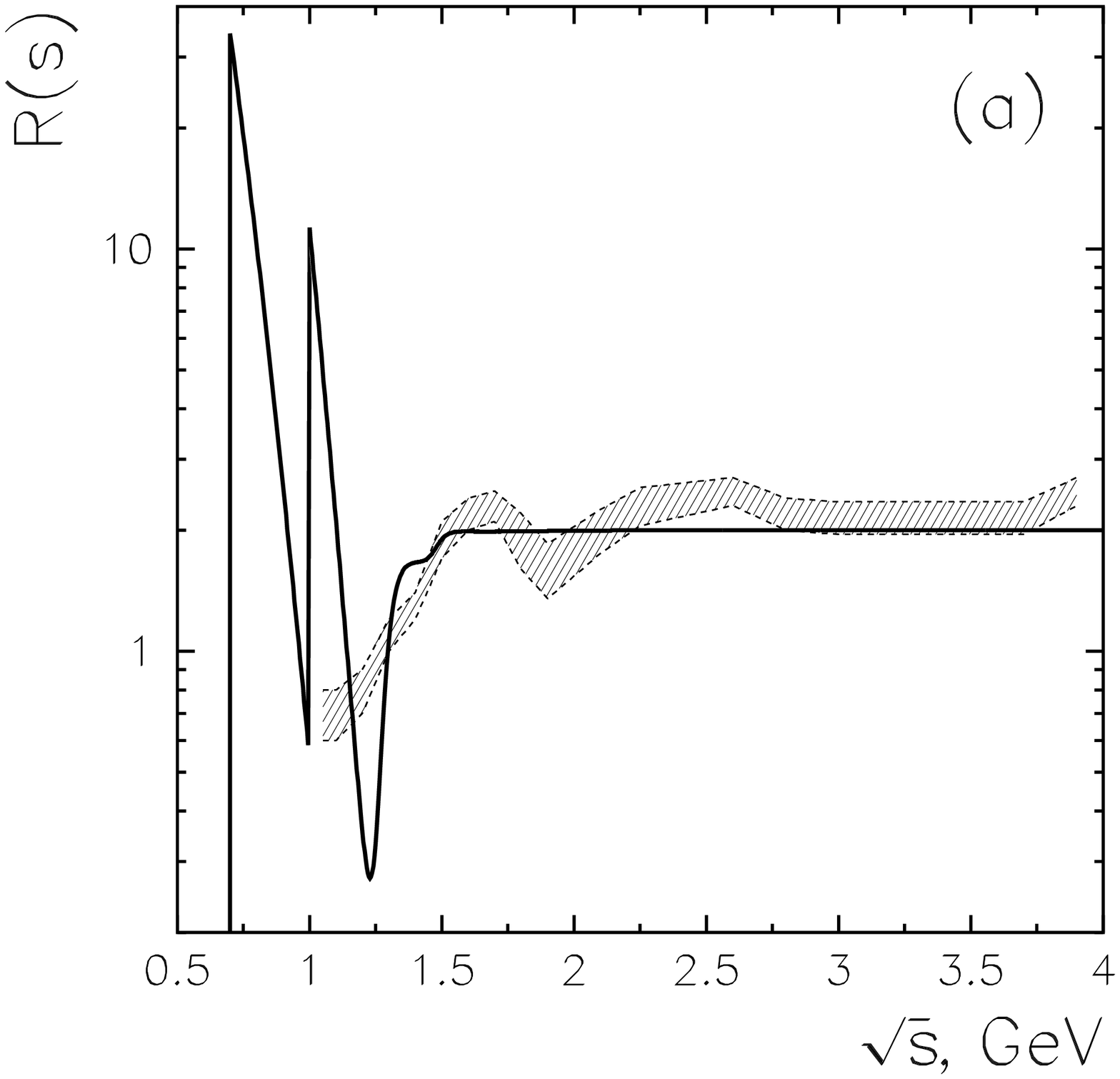,width=8cm}}
\vspace{1.0cm}
\centerline{\epsfig{file=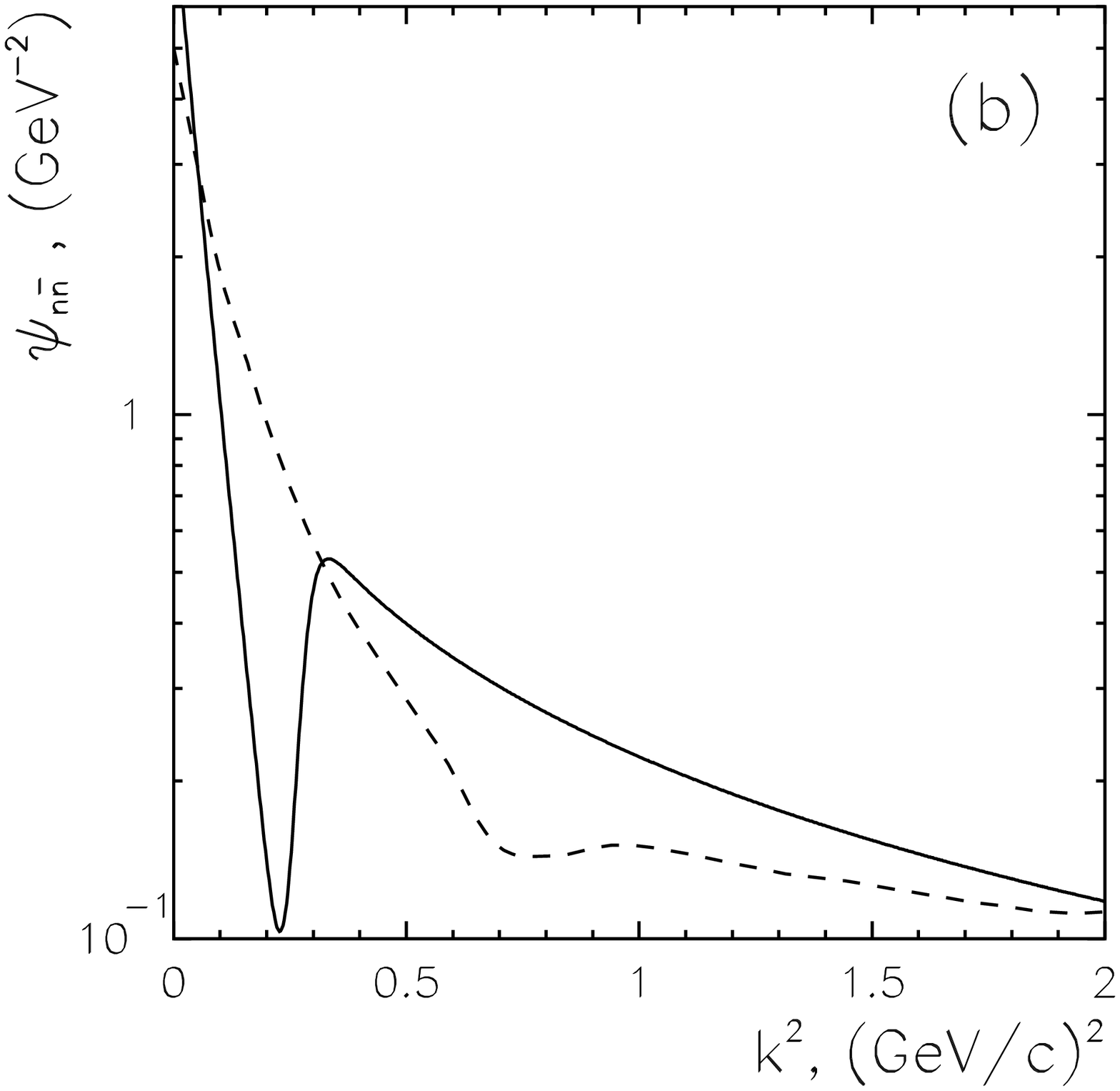,width=8cm}\hspace{1cm}
            \epsfig{file=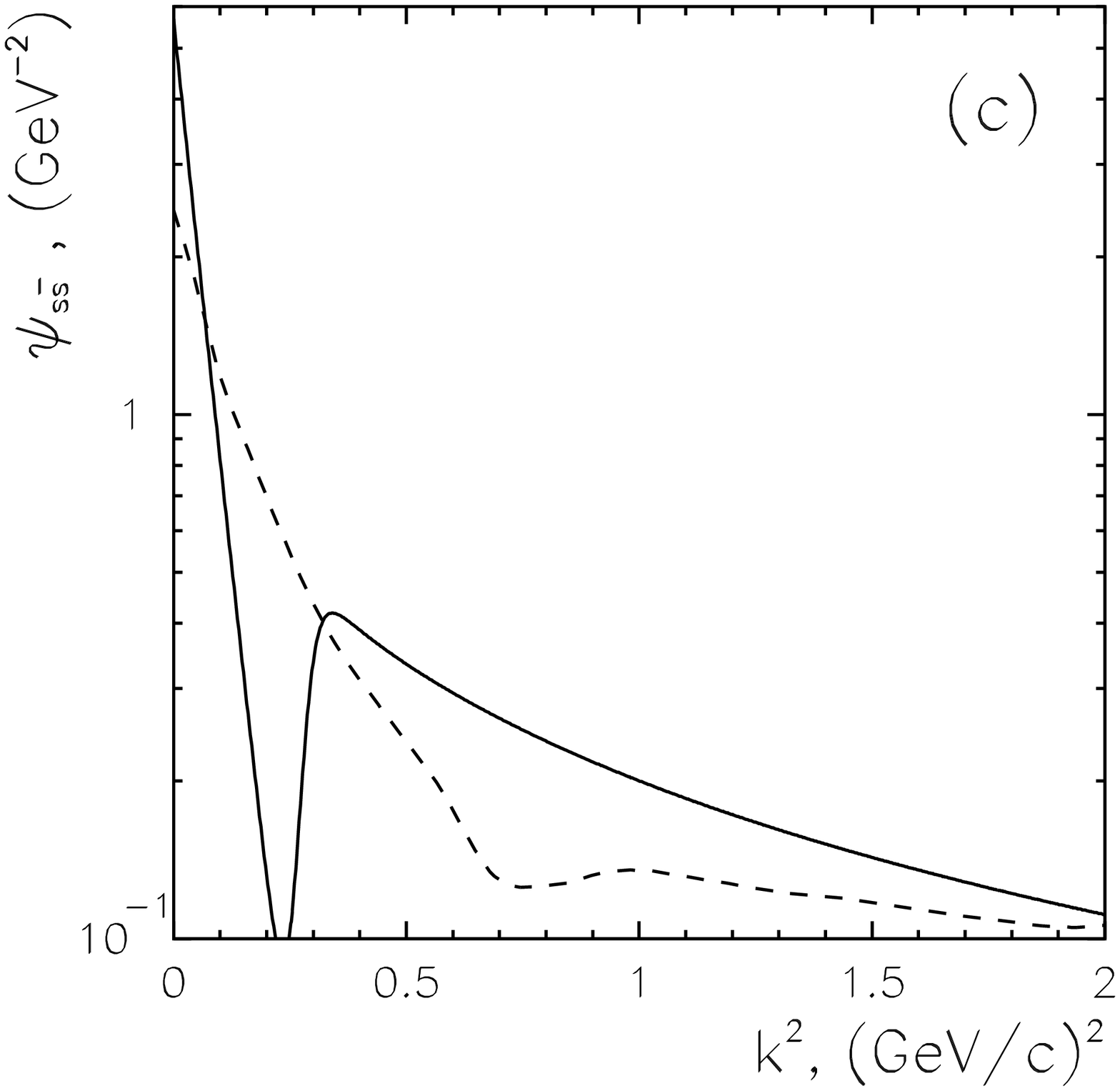,width=8cm}}
\caption{a) $R_{\rm vert}(s)$ (solid line, Eq. (40)) {\it vs}
$R(s)=\sigma( e^+e^-\to hadrons)/\sigma( \mu^+\mu^-\to hadrons)$
(hatched area); b,c)  $k^2$-dependence of photon wave function
($k^2$ is relative quark momentum squared):
$\Psi_{\gamma\to n\bar n}(k^2)$ (b) and
$\Psi_{\gamma\to s\bar s}(k^2)$ (c). Solid curves stand for new wave
functions and dashed lines for old one. }
\end{figure}

\newpage

\begin{figure}
\centerline{\epsfig{file=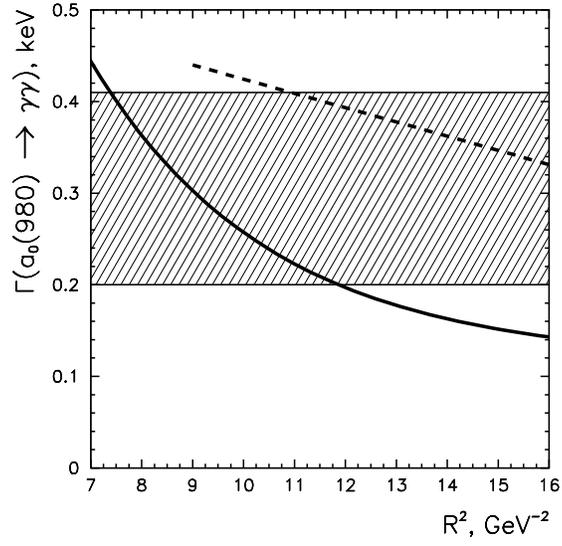,width=8cm}}
\caption{Partial width $\Gamma\left(a_0(980)\to \gamma\gamma\right)$
calculated under the assumption that $a_0(980)$ is $q\bar q$ system,
being a function of radius square of $a_0(980)$. Solid curve stands
for the calculation with new photon wave function, dotted curve
stands for old one. Shaded area corresponds to the values allowed by
the data [22].} \end{figure}

\newpage
\begin{figure}
\centerline{\epsfig{file=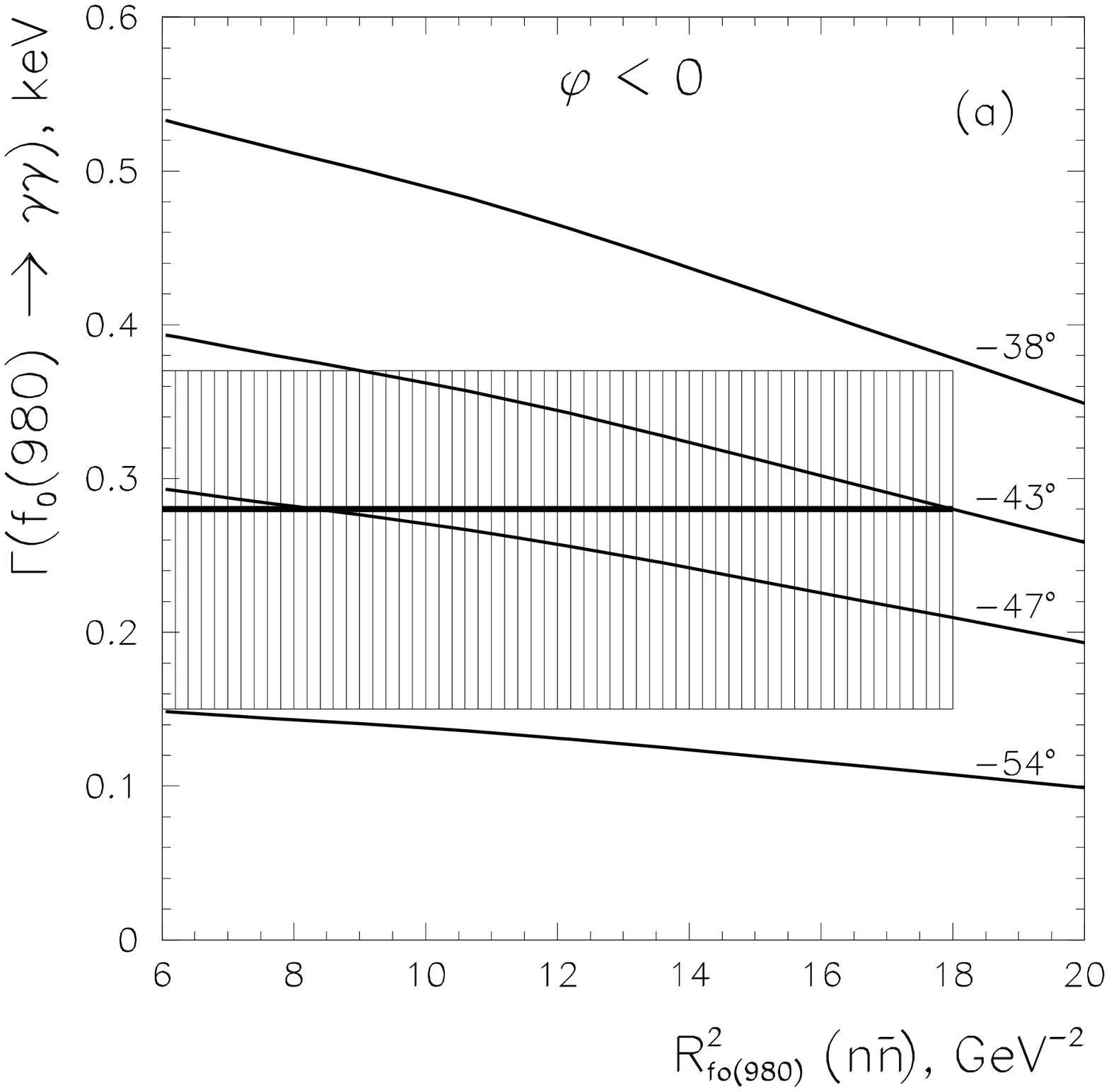,width=8cm}\hspace{1cm}
            \epsfig{file=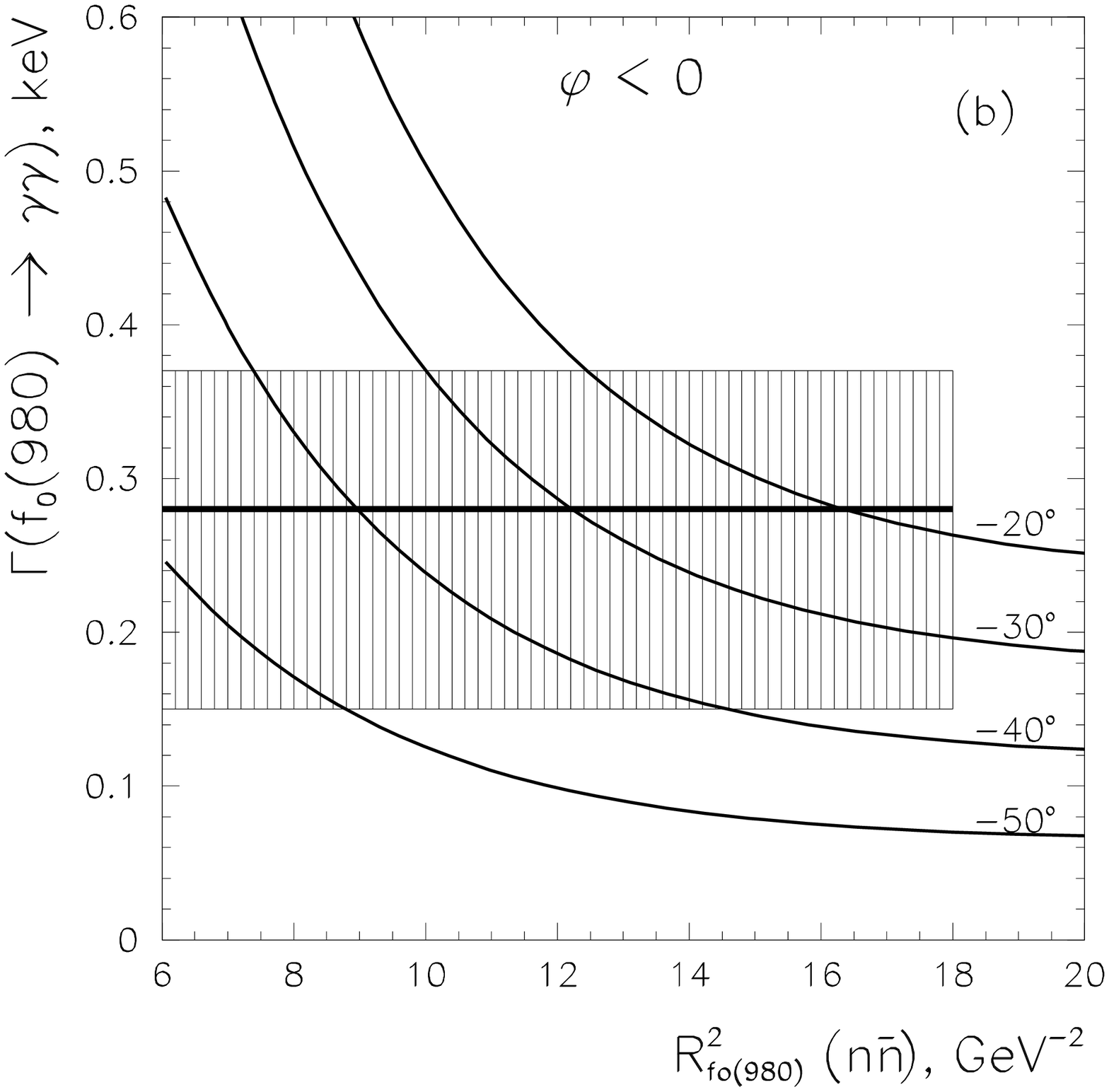,width=8cm}}
\caption{Partial width $\Gamma\left(f_0(980)\to \gamma\gamma\right)$
calculated under the assumption that $f_0(980)$ is $q\bar q$ system,
$q\bar q= n\bar n \cos\varphi+s\bar s\sin\varphi$, depending on radius
squared of the $q\bar q$ system: (a) with old photon wave function,
(b) with new one. Calculations were carried out with different values
of mixing angle $\varphi$ in the region $\varphi<0$.  Shaded area
 shows the allowed experimental values [26]. }
\end{figure}

\begin{figure}
\centerline{\epsfig{file=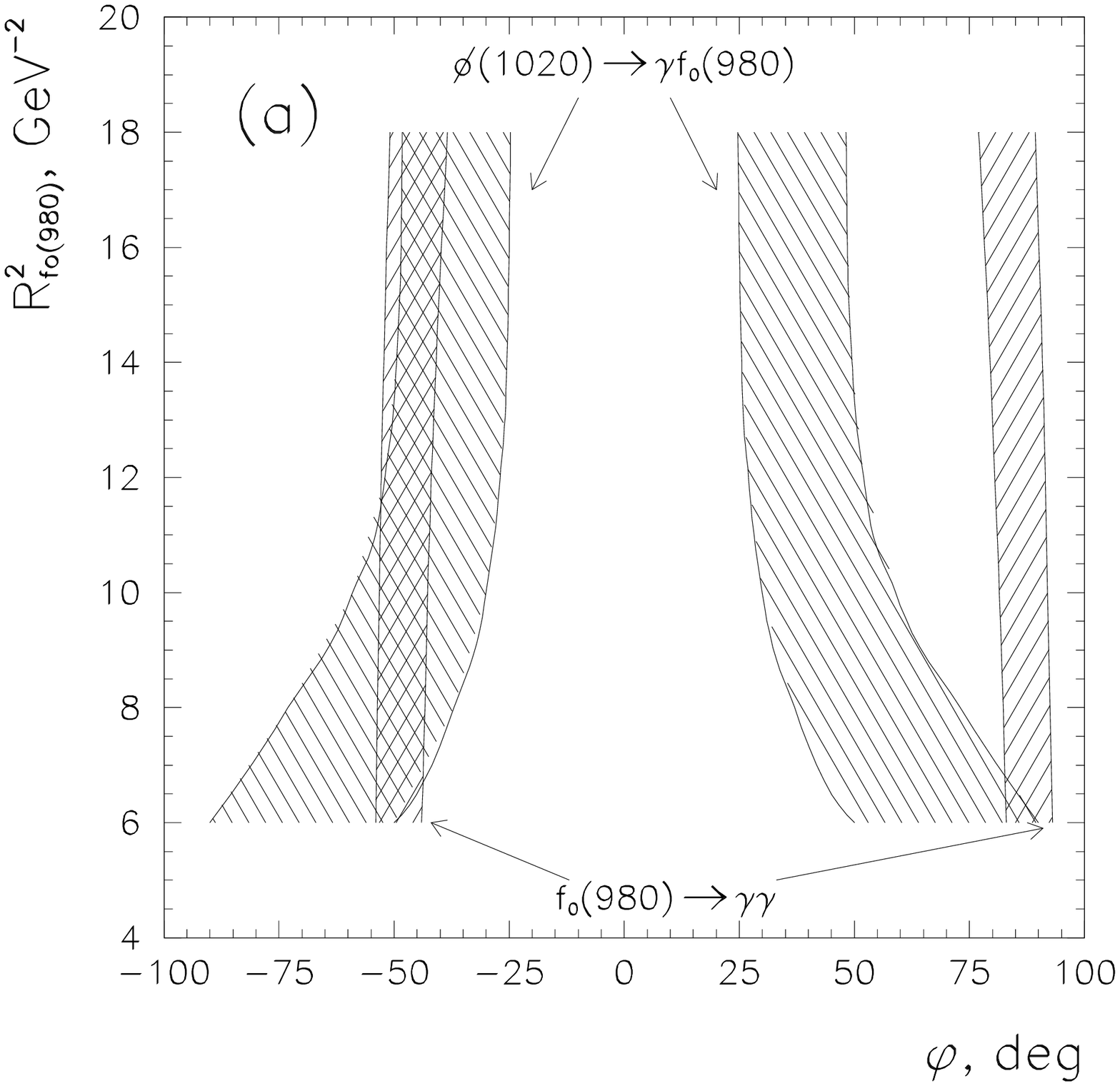,width=8cm}\hspace{1cm}
            \epsfig{file=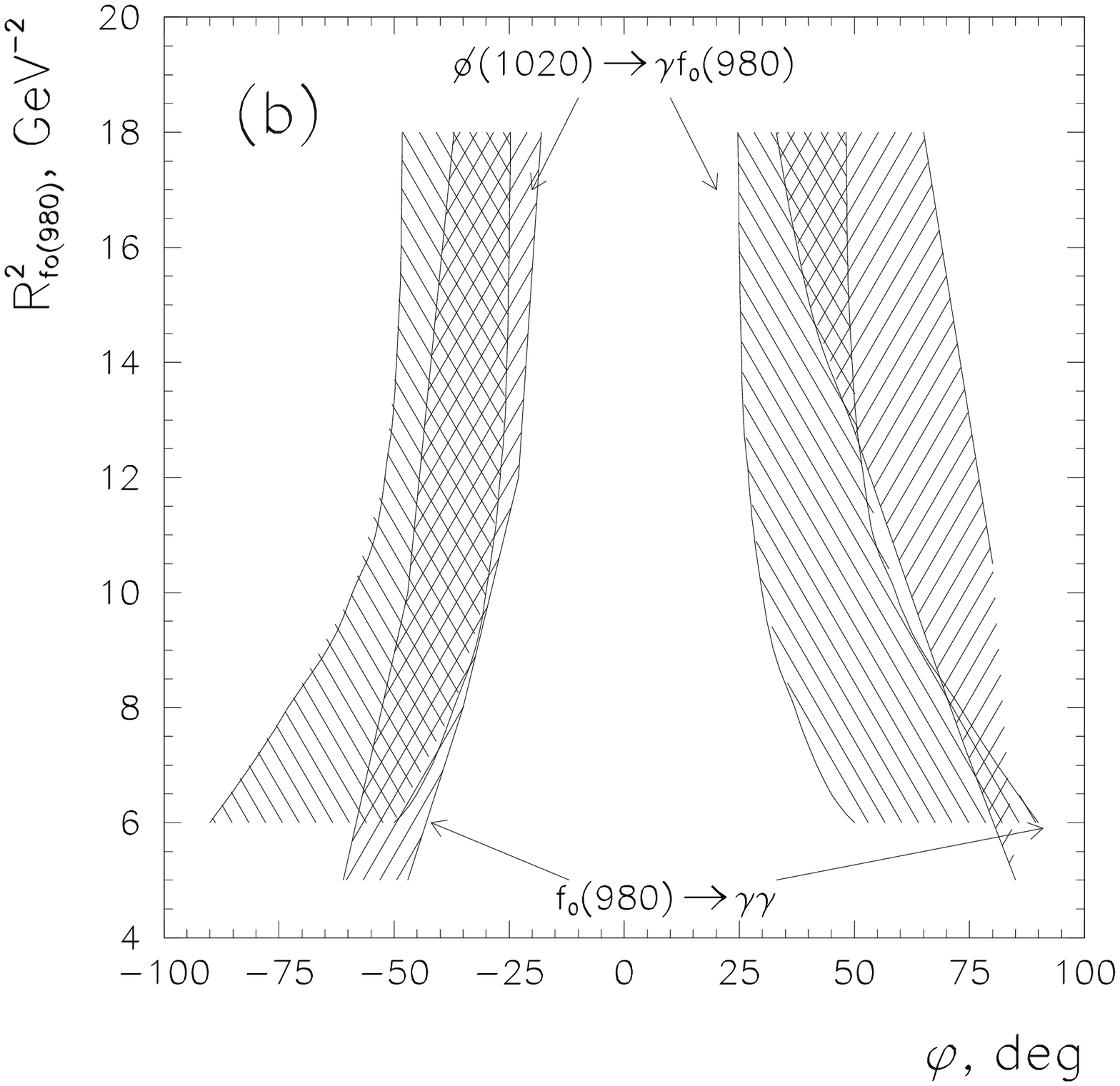,width=8cm}}
\caption{Combined presentation of the $(R^2_{f_0(980)}, \varphi)$ areas
allowed by the experiment
for the decays $f_0(980)\to \gamma\gamma$ and $\phi(1020)\to \gamma
f_0(980)$ with old (a) photon wave function and new one (b).}
\end{figure}

\begin{figure}
\centerline{\epsfig{file=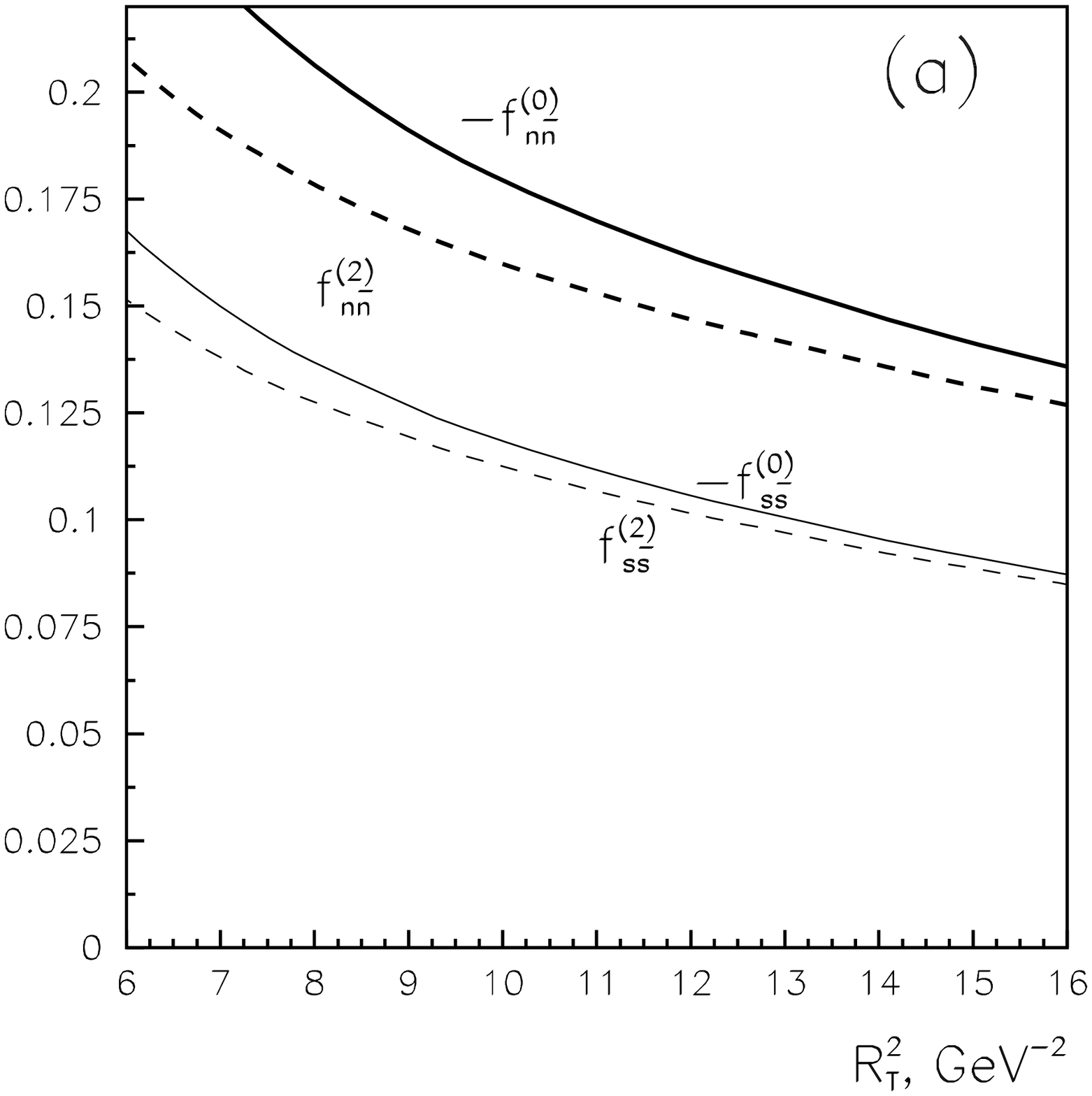,width=8cm}\hspace{1cm}
            \epsfig{file=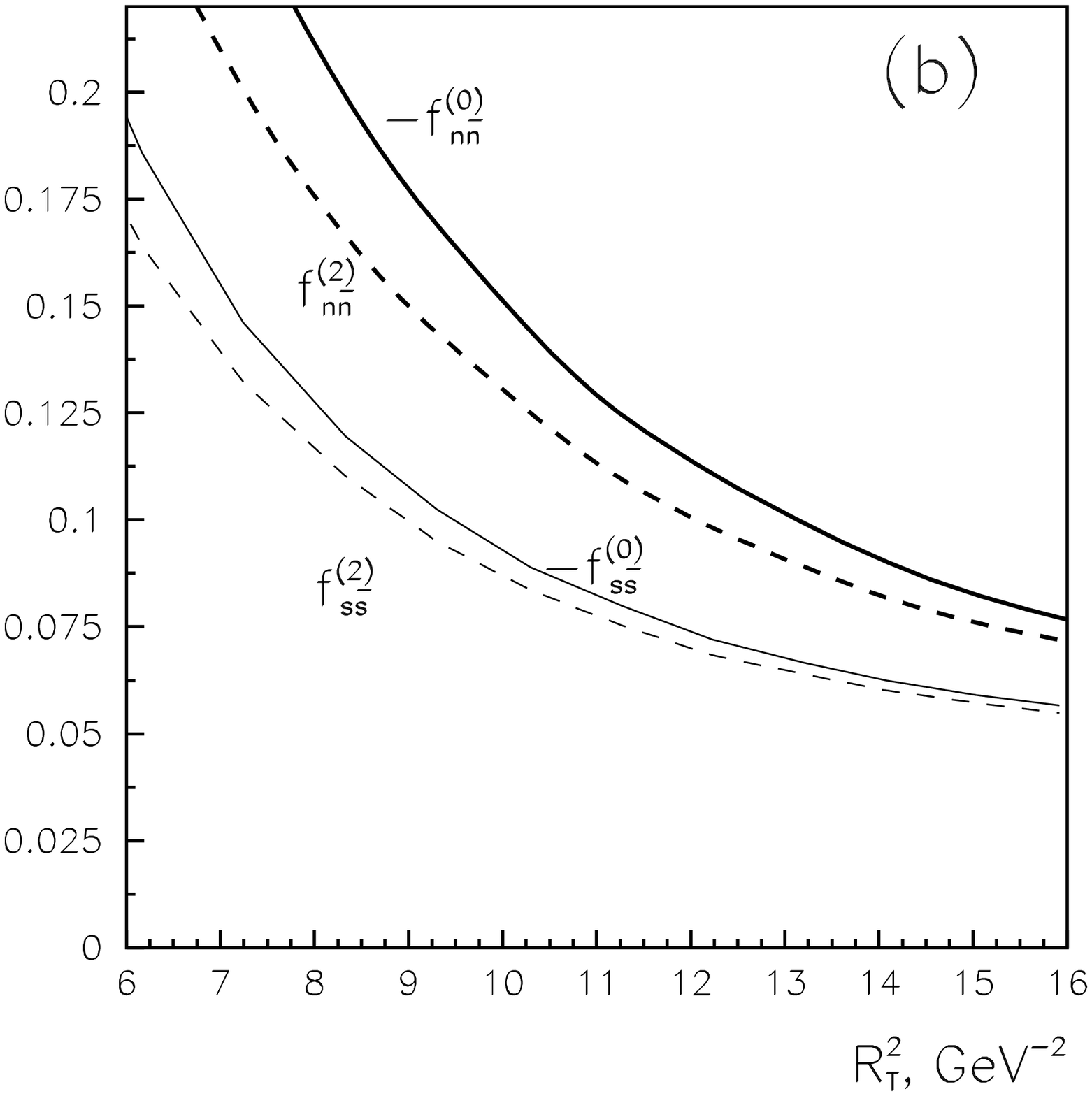,width=8cm}}
\caption{Transition form factors in the decay of tensor
quark--antiquark states $1^3P_2 n\bar n\to \gamma\gamma$ and
$1^3P_2 s\bar s\to \gamma\gamma$ as functions of radius squared of the
$q\bar q$ system calculated with old (a) and new (b) photon wave
functions.}
\end{figure}

\begin{figure}
\centerline{\epsfig{file=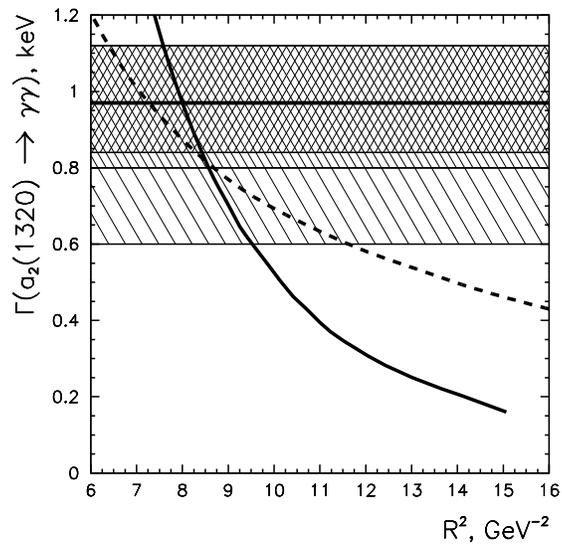,width=8cm}}
\caption{Calculated curves $vs$ experimental data (shaded areas) for
$\Gamma\left(a_2(1320)\to \gamma\gamma\right)$. Solid curve stands for
new photon wave function and dotted line for old one.}
\end{figure}

\begin{figure}
\centerline{\epsfig{file=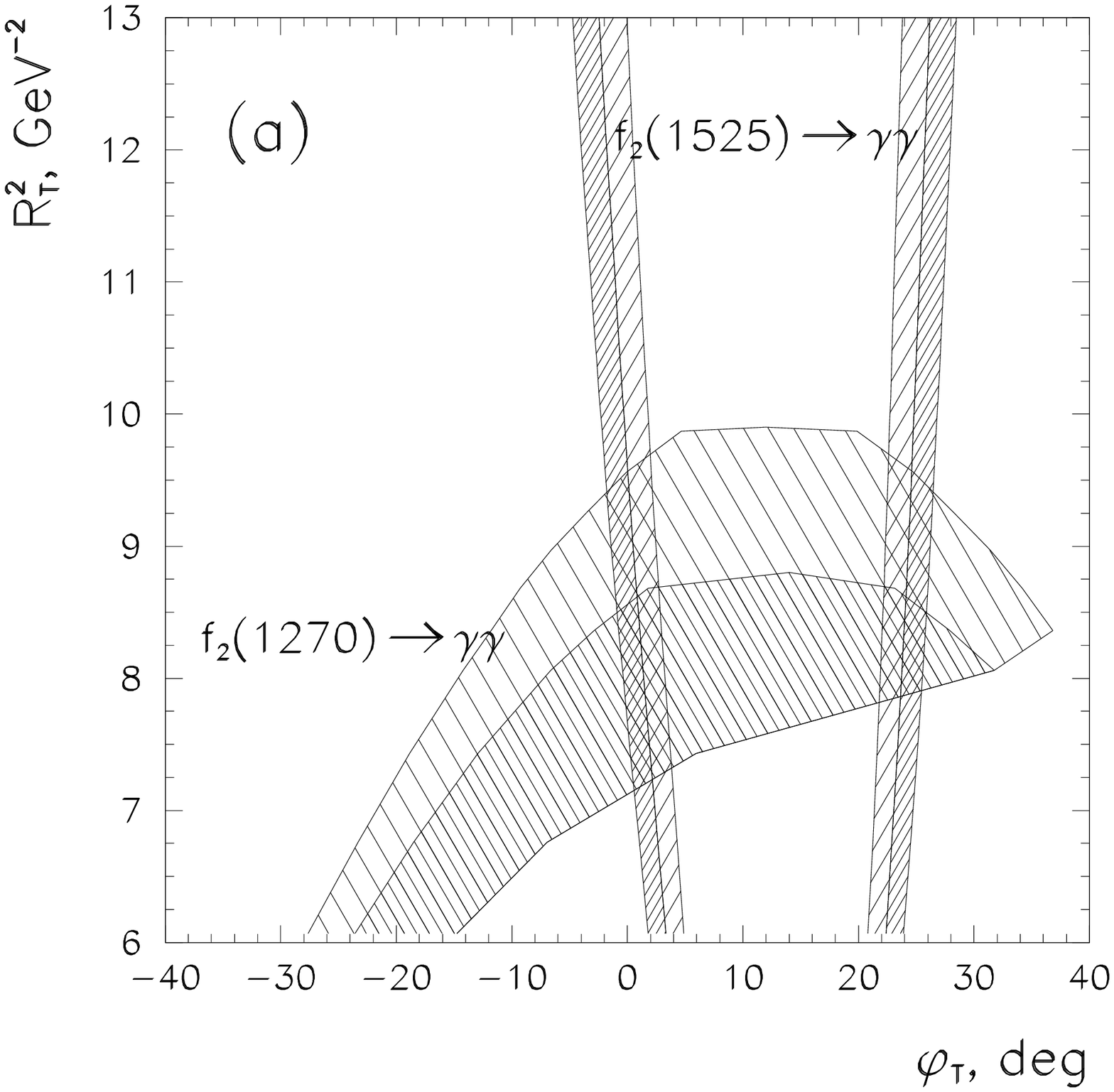,width=8cm}\hspace{1cm}
            \epsfig{file=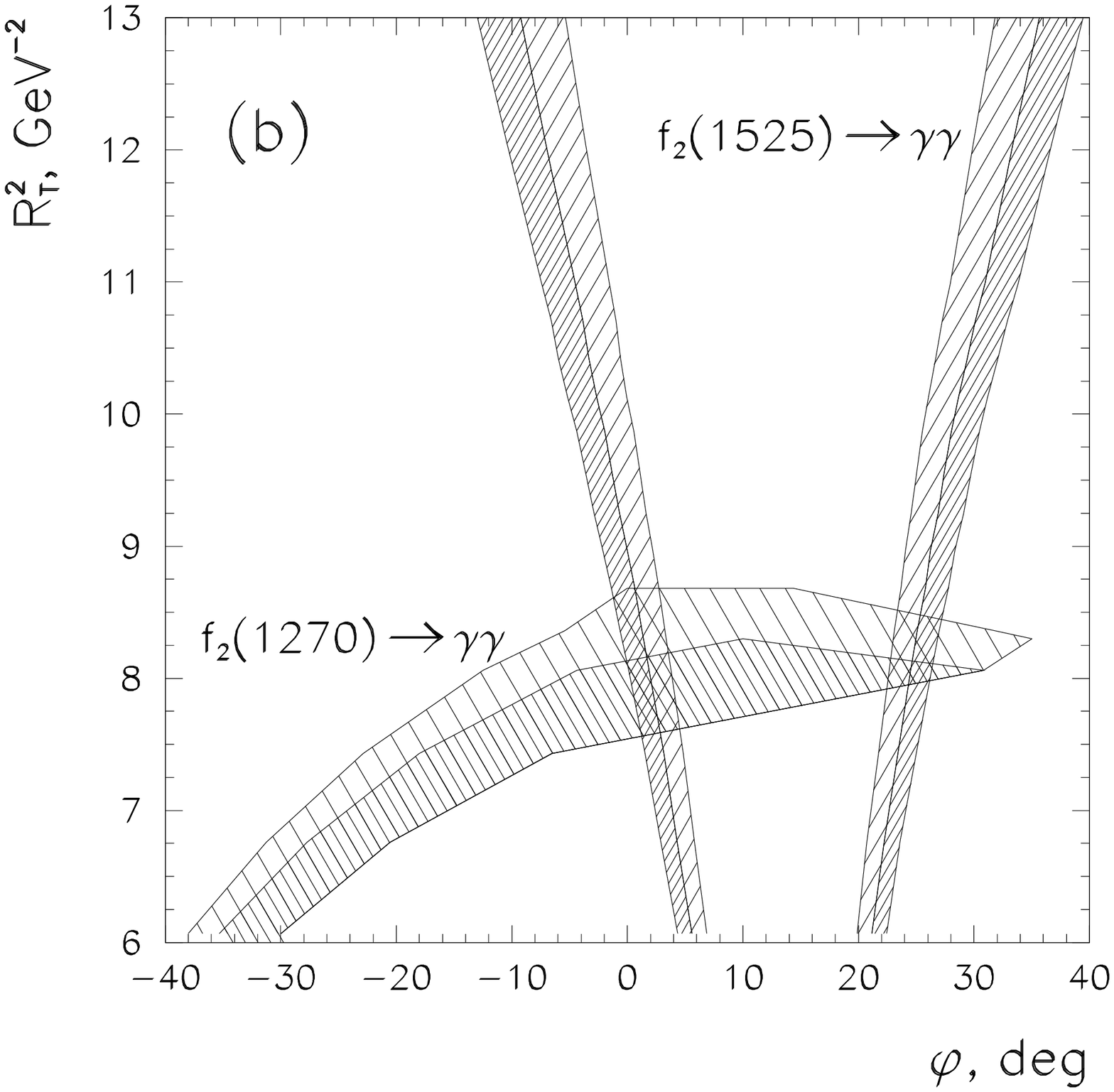,width=8cm}}
\caption{Allowed areas $(R^2_{f_0(980)}, \varphi)$ for partial widths
$\Gamma\left(f_2(1270)\to \gamma\gamma\right)$ and
$\Gamma\left(f_2(1525)\to \gamma\gamma\right)$ calculated with old (a)
and new (b) photon wave functions. Mixing angle $\varphi_T$ defines
flavour content of mesons as follows:
$f_2(1270)=n\bar n\cos \varphi_T+s\bar s \sin\varphi_T$ and
$f_2(1525)=-n\bar n\sin \varphi_T+s\bar s \cos\varphi_T$.}
\end{figure}

\end{document}